\documentclass[a4paper,11pt]{article}
\pdfoutput=1 % if your are submitting a pdflatex (i.e. if you have
\usepackage{jheppub}
\usepackage{cite} % for details on the use of the package, please
\allowdisplaybreaks

\title{Spiky strings in de Sitter space} 
\author[a]{Mitsuhiro Kato,}

\author[b]{Kanji Nishii,}

\author[b]{Toshifumi Noumi,}

\author[b]{Toshiaki Takeuchi,}

\author[c]{Siyi Zhou}

\affiliation[a]{Institute of Physics, University of Tokyo, Komaba, Meguro-ku, Tokyo 153-8902, Japan}
\affiliation[b]{Department of Physics, Kobe University, Kobe 657-8501, Japan}
\affiliation[c]{The Oskar Klein Centre for Cosmoparticle Physics \& Department of Physics, Stockholm University,
	AlbaNova, 106 91 Stockholm, Sweden.}

\emailAdd{kato@hep1.c.u-tokyo.ac.jp}

\emailAdd{kanji.nishii@stu.kobe-u.ac.jp}

\emailAdd{tnoumi@phys.sci.kobe-u.ac.jp}

\emailAdd{toshi.takeuchi@stu.kobe-u.ac.jp}

\emailAdd{siyi.zhou@fysik.su.se}

\preprint{UT-Komaba/21-1, KOBE-COSMO-21-04}

\abstract{
We study semiclassical spiky strings in de Sitter space and the corresponding Regge trajectories, generalizing the analysis in anti-de Sitter space. In particular we demonstrate that each Regge trajectory has a maximum spin due to de Sitter acceleration, similarly to the folded string studied earlier. While this property is useful for the spectrum to satisfy the Higuchi bound, it makes a nontrivial question how to maintain mildness of high-energy string scattering which we are familiar with in flat space and anti-de Sitter space. Our analysis implies that in order to have infinitely many higher spin states, one needs to consider infinitely many Regge trajectories with an increasing folding number.}

\begin{document} 
\setcounter{tocdepth}{2}
\maketitle
\flushbottom

\section{Introduction}
\setcounter{equation}{0}

Understanding de Sitter space in string theory is a challenging, but definitely important issue. In the past decades, intensive efforts have been made toward construction of de Sitter space in string theory (see, e.g., Ref.~\cite{Danielsson:2018ztv} for a review). By evading the assumptions of the Maldacena-Nunez no-go theorem~\cite{Maldacena:2000mw}, several promising scenarios such as the KKLT scenario~\cite{Kachru:2003aw} and the Large Volume Scenario~\cite{Conlon:2005ki} have been proposed. While these scenarios are based on ingredients that are individually justifiable, it is still an ongoing issue whether one can write down an explicit and fully-controlled compactification that unifies all the necessary components. There are also some attempts~\cite{Obied:2018sgi,Garg:2018reu,Ooguri:2018wrx} in the swampland program which try to interpret the nontriviality as an obstruction to de Sitter space in string theory. Thus it is tempting to search for complementary approaches to this journey. 

\medskip
This paper is a continuation of the previous work~\cite{Noumi:2019ohm} by three of the present authors, which initiated such a complementary approach  from the worldsheet theory perspective. More specifically, we study the semiclassical spectrum of would-be string worldsheet theory in de Sitter space, utilizing developments on integrability in the AdS/CFT correspondence. Semiclassical spectra of worldsheet theory in various curved spacetimes have been studied since seminal works by de Vega and Sanchez in 80's~\cite{deVega:1987veo,deVega:1988jh} and the followups~\cite{Combes:1993rw,deVega:1994yz,deVega:1996mv}. Researches in this direction have been further boosted with the advent of the AdS/CFT correspondence~\cite{Maldacena:1997re}, especially since the Gubser-Klebanov-Polyakov analysis~\cite{Gubser:2002tv} of folded strings~\cite{deVega:1996mv}. As nicely reviewed in Ref.~\cite{Tseytlin:2010jv}, various semiclassical solutions in AdS were then constructed and studied by using the integrability technique~\cite{Berenstein:2002jq,Frolov:2002av,Minahan:2002rc,Frolov:2003qc,Frolov:2003xy,Engquist:2003rn,Arutyunov:2003za,Larsen:2003tb,Ryang:2004tq,Kruczenski:2004cn,Kruczenski:2004wg,Ryang:2005yd,Park:2005kt,Hofman:2006xt,Dorey:2006dq,McLoughlin:2006tz,Chen:2006gea,Arutyunov:2006gs,Minahan:2006bd,Spradlin:2006wk,Bobev:2006fg,Kruczenski:2006pk,Okamura:2006zv,Ryang:2006yq,Ishizeki:2007we,Mosaffa:2007ty,Hayashi:2007bq,Ishizeki:2007kh,Mikhailov:2007xr,Jevicki:2007aa,Kruczenski:2008bs,Jevicki:2008mm,Miramontes:2008wt,Ryang:2008hr,Abbott:2008yp,Ishizeki:2008tx,Hollowood:2009tw,Jevicki:2009uz,Tirziu:2009ed,Kruczenski:2010xs}. Furthermore, progress in the past several years reaches beyond the string spectrum to include holographic higher-point correlation functions~\cite{Zarembo:2010rr,Costa:2010rz,Janik:2011bd,Kazama:2011cp,Caetano:2012ac,Kazama:2013qsa,Kazama:2016cfl}. These developments motivate us to perform a similar analysis in de Sitter space toward understanding of de Sitter space in string theory.

\medskip
In Ref.~\cite{Noumi:2019ohm}, as a first step in this direction, we revisited the folded string spectrum in de Sitter space~\cite{deVega:1996mv} and studied its consistency with the Higuchi bound~\cite{Higuchi:1986py}, a unitarity bound on the mass of higher-spin particles in de Sitter space. For a bosonic higher-spin particle with the mass $m$ and the spin $s$, the Higuchi bound reads $m^2\geq s(s-1)H^2$, where $H$ is the Hubble scale, so that a naive extrapolation of the flat space Regge trajectory violates the bound at a high energy scale.
Then, one might wonder if the Higuchi bound could be an obstruction to de Sitter space in string theory. However, a careful study of the trajectory is needed since the potential violation is at the horizon scale~\cite{Noumi:2019ohm,Lust:2019lmq}.
In Ref.~\cite{Noumi:2019ohm}, we explicitly demonstrated that the Regge trajectory is modified by the curvature effects appropriately such that the Higuchi bound is satisfied. In particular, in sharp contrast to flat space and AdS, each Regge trajectory in de Sitter space has a maximum spin because of causality and existence of the cosmological horizon. While this property is crucial to evade a potential conflict with the Higuchi bound, it makes a nontrivial question how to maintain mildness of high-energy string scattering which we are familiar with in flat space and AdS\footnote{A possible scenario to avoid this subtlety is to postulate that the higher-spin spectrum is the same as the flat space one up to the Planck scale, which provides an interesting bound on the inflation scale~\cite{Noumi:2019ohm,Lust:2019lmq}.}. It is thus intriguing to explore this direction further, by considering different types of semiclassical string solutions and the corresponding Regge trajectories. In this paper, we study spiky strings in de Sitter space, generalizing the analysis in AdS~\cite{Kruczenski:2004wg,Frolov:2002av}.

\medskip
The organization of the paper is as follows: In Sec.~\ref{semiclassicaldS}, we review basics of worldsheet theory in de Sitter space. In Sec.~\ref{foldedstrings}, we study folded strings with internal motion. Then, in Sec.~\ref{spikystrings} and Sec.~\ref{stringinternal}, we study more general spiky string solutions. We conclude in Sec.~\ref{conclusion} with discussion of our results. Some technical details are given in Appendix.

\section{Setup}\label{semiclassicaldS}

In this section we summarize basics of the worldsheet theory in de Sitter space necessary for our semiclassical analysis. See also Ref.~\cite{Tseytlin:2010jv} for a nice review on semiclassical strings in AdS. Our argument is analogous to the one there except for the fact that de Sitter space has an acceleration and a cosmological horizon accordingly, which turns out to bring about qualitative differences from the flat space and AdS cases.

\subsection{Target space}

\begin{figure}[t] 
	\centering 
	\includegraphics[width=11cm]{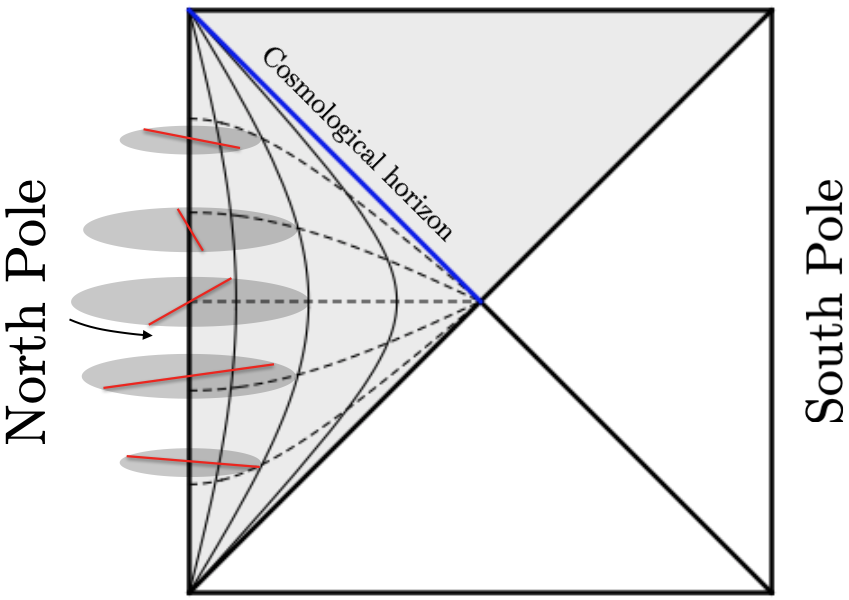}  
	\caption{Penrose diagram of $dS_3$:
Each point represents an $S^1$ subspace and the edges correspond to the north and south poles of $S^{2}$. For example, the planar coordinates cover a half of the whole space (the shaded region) and the cosmological horizon for an observer sitting at the north pole is  the blue line. The static coordinates cover a half of the planar coordinates $Y_3 \pm Y_0 \geq 0$, that is inside the cosmological horizon.
The dotted and rigid curves are sections of $\text{constant } t \text{ and } {\rm r}(=\sin\rho),$ respectively. We study strings rotating around the center $\rm r=0$ ($\rho=0$) of the static coordinate. 
} \label{dS_penrose}
\end{figure}

In this paper we study string Regge trajectories on $dS_3\times S^1$ (which may also be identified with an appropriate subspace of a larger target space). The three-dimensional de Sitter space $dS_3$ with the radius $R$ is defined by a hypersurface,
\begin{align}
-Y_0^2+Y_1^2+Y_2^2+Y_3^2=R^2\,,
\end{align}
embedded into a four-dimensional Minkowski space with the line element,
\begin{align}
ds^2=-dY_0^2+dY_1^2+dY_2^2+dY_3^2\,.
\end{align}
For our purpose, it is convenient to use the static coordinates:
\begin{align}
\label{static_embedding}
\frac{Y_3+Y_0}{R}=\sqrt{1-{\rm r}^2\,} e^t\,,
\quad
\frac{Y_3-Y_0}{R}=\sqrt{1-{\rm r}^2\,} e^{-t}\,,
\quad
\frac{Y_1+iY_2}{R}={\rm r}\,e^{i\phi}\,,
\end{align}
where $-\infty <t<\infty$, $0\leq {\rm r}\leq 1$, and $\phi$ has a periodicity $2\pi$. The corresponding metric is
\begin{align}
ds^2=R^2\bigg[-(1-{\rm r}^2)dt^2+\frac{d{\rm r}^2}{1-{\rm r}^2}+{\rm r}^2d\phi^2\bigg]
\,.
\end{align}
As depicted in Fig.~\ref{dS_penrose}, this coordinate system covers a quarter of the full de Sitter space. An observer sitting at the origin ${\rm r}=0$ has a cosmological horizon at ${\rm r}=1$, hence this coordinate system can be used to describe the inside of the horizon. 

\medskip
To utilize results in AdS, it is convenient to introduce a coordinate $\rho$ defined by $\sin\rho={\rm r}$ ($0\leq\rho\leq\pi/2$), in terms of which the metric reads
\begin{align}
\label{static_metric}
ds^2=R^2\left(-\cos^2\!\rho \,dt^2+d\rho^2+\sin^2\rho\,d\phi^2
\right)
\,.
\end{align}
Note that in these coordinates, the observer sitting at the origin and the cosmological horizon are located at $\rho=0$ and $\rho=\pi/2$, respectively. Since global coordinates of AdS are obtained by a Wick rotation,
\begin{align}\label{wickrotationadstods}
\rho\to -i\rho\,,\quad
t\to it\,,
\quad
R^2\to -R^2\,,
\end{align}
we may generalize semiclassical solutions in AdS to de Sitter space in a straightforward manner.
Together with an internal $S^1$ parameterized by the coordinate $\varphi$, our target space metric is given by
\begin{align}
\label{static_extra}
ds^2=R^2\left(-\cos^2\!\rho \,dt^2+d\rho^2+\sin^2\rho\, d\phi^2+d\varphi^2\right)\,,
\end{align}
where for generality we leave the periodicity of $\varphi$ a free parameter. In other words, we absorb the radius of the circle $S^1$ into the definition of $\varphi$.

\subsection{Worldsheet theory}
\label{subsec:worldsheet}

Let us consider the Nambu-Goto string on the target space~\eqref{static_extra}:
\begin{align} \label{nambugoto}
S_{NG} &  
=-\frac{1}{2\pi\alpha' } \int d\tau d\sigma\sqrt{-\dot X^2 X'^2 + (\dot X \cdot X')^2 } \,,
\end{align}
where $(2\pi\alpha')^{-1}$ is the string tension and we defined
\begin{align}
\dot{X}^2=G_{AB}\dot{X}^A\dot{X}^B\,,
\quad
X'^2=G_{AB}X'^AX'^B\,,
\quad
\dot{X}\cdot X'=G_{AB}\dot{X}^AX'^B\,,
\end{align}
with $X^A=(t,\rho,\phi,\varphi)$ and a  target space metric,
\begin{align}
G_{AB}=R^2 \cdot{\rm diag}\left(-\cos^2\rho,1,\sin^2\rho,1\right)\,.
\end{align}
Also the dot and prime stand for derivatives in the worldsheet time coordinate $\tau$ and the worldsheet spatial coordinate $\sigma$, respectively. The equation of motion for $X^A$ reads
\begin{align} \nonumber 
	0 & = \partial_{\tau}\left[\frac{G_{A B}\left(\dot{X}^{B} X^{\prime 2}-X^{{\prime}B}\left(\dot{X} \cdot X^{\prime}\right)\right)}{\sqrt{-\dot{X}^{2} X^{\prime 2}+\left(\dot{X} \cdot X^{\prime}\right)^{2}}}\right]+\partial_{\sigma}\left[\frac{G_{A B}\left(X^{{\prime}B} \dot{X}^{2}-\dot{X}^{B}\left(\dot{X} \cdot X^{\prime}\right)\right)}{\sqrt{-\dot{X}^{2} X^{\prime 2}+\left(\dot{X} \cdot X^{\prime}\right)^{2}}}\right] \\ \label{eomconstant}
	&\quad
	- \frac{\partial_{A} G_{B C}\left[\dot{X}^{B} \dot{X}^{C} X^{\prime 2}+X^{{\prime}B} X^{{\prime}C} \dot{X}^{2}-2 \dot{X}^{B} X^{{\prime}C}\left(\dot{X} \cdot X^{\prime}\right)\right]}{2 \sqrt{-\dot{X}^{2} X^{\prime 2}+\left(\dot{X} \cdot X^{\prime}\right)^{2}}} \,.
\end{align}

\paragraph{Rigid string ansatz.}

Classical string solutions discussed in this paper are captured by the following ansatz for closed string configurations:
\begin{align}
\label{rigid_ansatz}
t=\tau\,,
\quad
\rho=\rho(\sigma)\,,
\quad
\phi=\omega \tau+N\sigma\,,
\quad
\varphi=\nu \tau
+\psi(\sigma)
\,,
\end{align}
where $\sigma$ has a periodicity $2\pi$ and we require $\rho(\sigma+2\pi)=\rho(\sigma)$ and $\psi(\sigma+2\pi)=\psi(\sigma)$,  assuming that the string has no winding along the circle $S^1$.  Also, $\omega$ and $\nu$ are constant angular velocities, and $N$ is an integer characterizing the ``winding" number along the angle $\phi$. Note that the case without internal space is covered simply by setting $\nu=\psi=0$. As depicted, e.g., in Fig.~\ref{threequantities}, the string at a fixed time $t=\tau$ is spreading on the two-dimensional $(\rho,\phi)$ plane. It then rotates along $\phi$ and $\varphi$ with angular velocities $\omega$ and $\nu$.

\medskip
With the ansatz~\eqref{rigid_ansatz}, the equations of motion~\eqref{eomconstant} reduce to the following (generally) independent three equations:
\begin{align}
0&=-\partial_\sigma\left[\frac{\rho'(\cos^2\rho-\omega^2\sin^2\rho-\nu^2)}{\sqrt{\mathcal{D}}}\right]
\nonumber
\\
\label{eom_rho}
&\quad
+\frac{1}{2}\frac{\sin2\rho[-(1+\omega^2)(\rho'^2+\psi'^2)+2N\nu\omega\psi'+N^2\cos2\rho-N^2\nu^2]}{\sqrt{\mathcal{D}}}\,,
\\
\label{eom_t}
0&=\partial_\sigma
\left[
\frac{\cos^2\rho(N\omega\sin^2\rho+\nu\psi')}{\sqrt{\mathcal{D}}}
\right]
\,,
\\
\label{eom_phi}
0&=\partial_\sigma
\left[
\frac{\nu\omega\sin^2\rho\,\psi'+N(\cos^2\rho-\nu^2)\sin^2\rho}{\sqrt{\mathcal{D}}}
\right]
\,,
\end{align}
where we introduced
\begin{align}
\mathcal{D}&=\frac{-\dot{X}^2X'^2+(\dot{X}\cdot X')^2}{R^4}
\nonumber
\\
&=(\cos^2\rho-\omega^2\sin^2\rho-\nu^2)\rho'^2
+(\cos^2\rho-\omega^2\sin^2\rho)\psi'^2
\nonumber
\\
&\quad
+2N\nu\omega\sin^2\rho\,\psi'
+N^2(\cos^2\rho-\nu^2)\sin^2\rho\,. \label{D}
\end{align}
Note that reality conditions require $\mathcal{D}\geq0$, otherwise the corresponding Nambu-Goto action becomes imaginary. Also one may show that when both $\mathcal{D}\neq 0$ and $\rho'\neq 0$ are satisfied, Eq.~\eqref{eom_rho} follows from Eqs.~\eqref{eom_t} and~\eqref{eom_phi}.

\paragraph{Energy, spin and internal $U(1)$ charge.}

To close the section, let us write down the energy $E$, spin $S$, and internal $U(1)$ charge $J$, which are of interest in the discussion of the Regge trajectory. Defining them as conjugates of $R\,t$, $-\phi$, and $-\varphi$, respectively, we have
\begin{align}\label{E_master}
E&= \frac{R}{2\pi\alpha'} \int_{0}^{2\pi} d \sigma \frac{ \cos^2 \rho (\rho'^2 +N^2 \sin^2 \rho+\psi'^2)  }{ \sqrt{\mathcal{D}}} \,, 
\\\label{S_master}
S&= \frac{R^2}{2\pi \alpha' } \int_0^{2\pi} d\sigma\frac{\sin^2\rho(\omega\rho'^2+\omega\psi'^2-N\nu\psi')}{\sqrt{\mathcal{D}}}\,,
\\\label{J_master}
J&= \frac{R^2}{2\pi\alpha'} \int_{0}^{2\pi} d \sigma  \frac{\nu\rho'^2+N^2\nu\sin^2\rho-N\omega\sin^2\rho\,\psi'}{\sqrt{\mathcal{D}}}\,,
\end{align}
which satisfies the following relation:
\begin{align}
\frac{R^2}{2\pi\alpha'} \int_{0}^{2\pi} d \sigma\sqrt{\mathcal{D}}
=RE-\omega S-\nu J\,. \label{worldsh_Ham}
\end{align}

\section{Folded strings}\label{foldedstrings}

\begin{figure}[t] 
	\centering 
	\includegraphics[width=14cm]{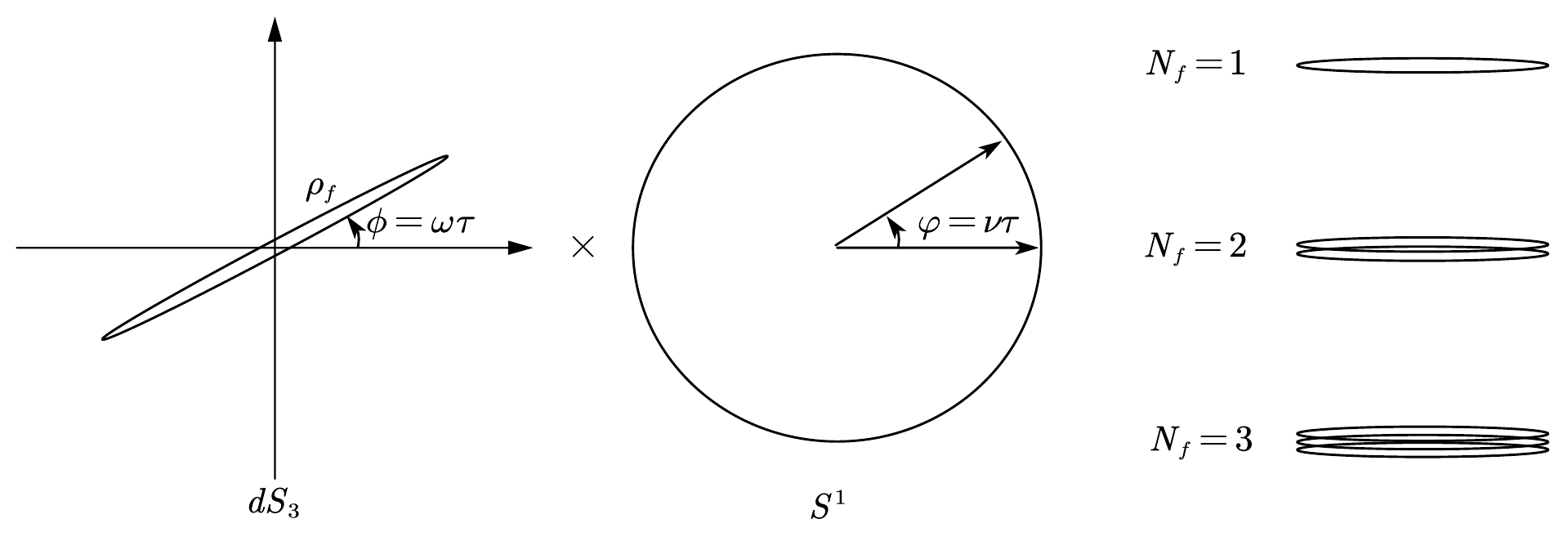}    
	\caption{Illustration of $\omega$, $\nu$ and $N_f$: $\omega$ is the angular velocity in the $\phi$ direction. $\nu$ is the angular velocity in the $\varphi$ direction. $N_f$ is the folding number. } \label{threequantities}
\end{figure}

We begin by generalizing our pervious work~\cite{Noumi:2019ohm} on rotating folded strings to include motion along the internal circle $S^1$. The folded string configuration is captured by the ansatz~\eqref{rigid_ansatz} with $N=\psi=0$, under which Eqs.~\eqref{eom_t}-\eqref{eom_phi} become trivial, whereas Eq.~\eqref{eom_rho} gives
\begin{align}
\partial_\sigma\left(\frac{\rho'}{|\rho'|}\right)
\sqrt{\cos^2\rho-\omega^2\sin^2\rho-\nu^2}=0
\,\,
\leftrightarrow
\,\,
\delta(\sigma-\sigma_{f})\sqrt{\cos^2\rho-\omega^2\sin^2\rho-\nu^2}=0\,.
\end{align}
Notice that the equation of motion is localized at the folding point $\sigma=\sigma_{f}$ where $\rho'$ flips the sign, simply because changes in the bulk profile $\rho(\sigma)$ ($\sigma\neq\sigma_{f}$) are gauge degrees of freedom associated to string reparameterization. Also the folding point satisfies
\begin{align}
\cos^2\rho-\omega^2\sin^2\rho-\nu^2=0\,,
\end{align}
and so it propagates with the speed of light, which is essentially the same as the familiar statement that open string end points propagate with the speed of light. Then, for given $\omega$ and $\nu$, the radius $\rho_{f}$ of the folding point is determined by
\begin{align}
\label{rho_f}
	\cot^2 \rho_{f} = \frac{\omega^2+\nu^2}{1-\nu^2}\,,
\end{align}
which is the maximum distance dictated by causality prohibiting superluminal propagation of the string. In general, closed strings may have multiple foldings, so that the solutions are parameterized by the angular velocities $\omega$ and $\nu$, and the folding number $N_{f}$. See Fig.~\ref{threequantities}.

\paragraph{Conserved charges.}

For these folded strings, the conserved charges~\eqref{E_master}-\eqref{J_master} read
\begin{align}\label{energyinternal}
	E & =\frac{1}{\sqrt{1-\nu^2}}\times\frac{4N_{f}R}{2\pi \alpha'} \int_{0}^{\rho_{f}} d\rho \frac{\cos^2 \rho}{\sqrt{1 - (\sin^2\rho/\sin^2\rho_{f}) }}
	~ , \\* \label{spininternal}
	S &  =\frac{\omega}{\sqrt{1-\nu^2}}\times\frac{4N_{f}R^2}{2\pi \alpha'} \int_{0}^{\rho_{f}} d\rho \frac{\sin^2 \rho}{\sqrt{1 - (\sin^2\rho/\sin^2\rho_{f})}}
~,\\* \label{chargeinternal}
	J &  =\frac{\nu}{\sqrt{1-\nu^2}}\times\frac{4N_{f}R^2}{2\pi \alpha'} \int_{0}^{\rho_{f}} d\rho \frac{1}{\sqrt{1 - (\sin^2\rho/\sin^2\rho_{f})}}
	~.
\end{align}
One may also rewrite them in terms of incomplete elliptic integrals,
\begin{align}
\mathcal{E}\left(\zeta | k^{2}\right) =\int_{0}^{\zeta} d \theta \sqrt{1-k^{2} \sin ^{2} \theta}~, \quad \mathcal{F}\left(\zeta | k^{2}\right) &=\int_{0}^{\zeta} d \theta \frac{1}{\sqrt{1-k^{2} \sin ^{2} \theta}}  ~,
\end{align}
as follows:
\begin{align}\label{E_folded_elliptic}
	E & =\frac{1}{\sqrt{1-\nu^2}}\times\frac{4N_{f}R}{2\pi \alpha'} \left[\sin^2\rho_{f}\,\mathcal{E}\left(\rho_{f} | \csc ^{2} \rho_{f}\right)+\cos^2\rho_{f}\,\mathcal{F}\left(\rho_{f} | \csc ^{2} \rho_{f}\right)\right]
	~ , \\ \label{S_folded_elliptic}
	S &  =\frac{\omega}{\sqrt{1-\nu^2}}\times\frac{4N_{f}R^2}{2\pi \alpha'}
\sin^2\rho_{f}
	\left[-\mathcal{E}\left(\rho_{f} | \csc ^{2} \rho_{f}\right)+\mathcal{F}\left(\rho_{f} | \csc ^{2} \rho_{f}\right)\right]
~,\\ \label{J_folded_elliptic}
	J &  =\frac{\nu}{\sqrt{1-\nu^2}}\times \frac{4N_{f}R^2}{2\pi \alpha'}  \mathcal{F}\left(\rho_{f} | \csc ^{2} \rho_{f}\right)\,.
\end{align} 
These expressions can be used to derive energy-spin relations and draw Regge trajectories.

\begin{figure}[t] 
	\centering 
	\includegraphics[width=7.3cm]{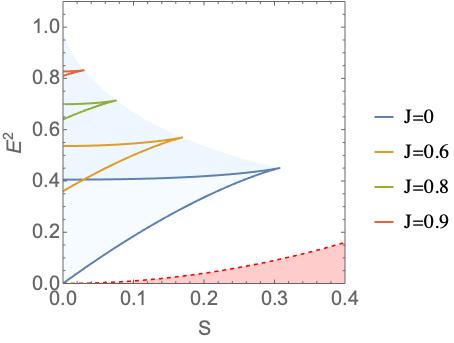}   \quad  
	\includegraphics[width=7.3cm]{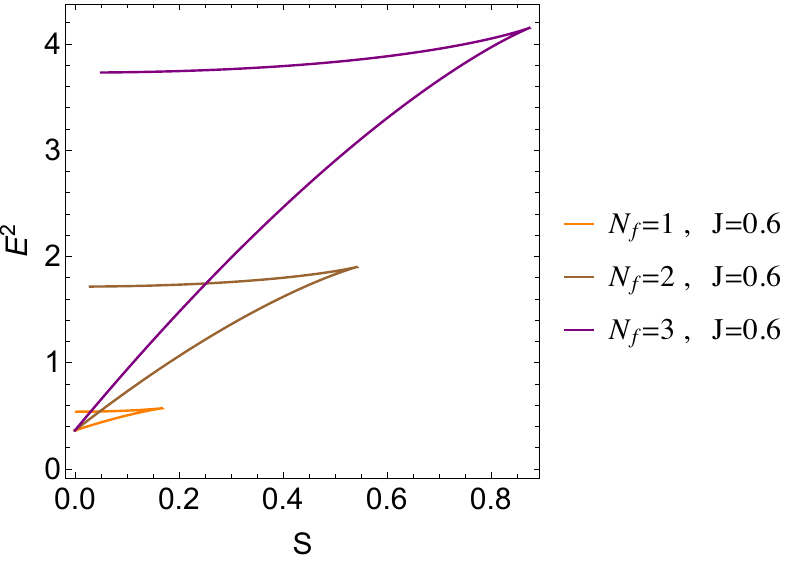}
	\caption{The left panel shows Regge trajectories for $N_{f} = 1$ with different internal charge $J$. The right panel shows Regge trajectories for different $N_{f}$ with $J = 0.6 R^2/\alpha^\prime$. The energy $E$, spin $S$ and internal charge $J$ are in the units of $R/\alpha'$, $R^2/\alpha'$ and $R^2/\alpha^\prime$, respectively.  We find that the Regge trajectories always satisfy the Higuchi bound ($E^2 \geq \frac{S(S-1)}{R^2}$), which prohibits  the red region.} \label{folded_Regge}
\end{figure} 

\paragraph{Regge trajectories.}
The left panel of Fig.~\ref{folded_Regge} shows Regge trajectories of one-folded strings ($N_{f}=1$) with a fixed internal charge $J$. First, the trajectory for $J=0$ accommodates a turning point, where the string has a maximum spin. This is a consequence of causality and existence of the cosmological horizon, which is helpful to satisfy the Higuchi bound~\cite{Noumi:2019ohm}. Next, if one increases the internal charge $J$, the trajectory shifts upwards simply because the internal motion increases the energy. Also the maximum spin decreases, so that the maximum spin of one-folded strings is the one for the $J=0$ string. Then, one-folded strings scan a finite region in the energy-spin plane represented by the blue shaded region. In particular, one needs to consider multiple folded strings ($N_{f}=2,3,\ldots$) to have infinitely many higher spins (see the right panel of Fig.~\ref{folded_Regge}). Note that the Higuchi bound is satisfied in the entire region. In the rest of the section, we study several limits and provide more quantitative arguments.

\subsection{Bound on internal charge $J$}
Fig.~\ref{folded_Regge} implies that for a fixed folding number $N_{f}$, there exists a maximum value of the internal charge $J$. To see this more quantitatively, let us recall
\begin{align}
\frac{\nu}{\sqrt{1-\nu^2}}\leq \sqrt{\frac{\nu^2+\omega^2}{1-\nu^2}}=\cot \rho_{f}\,,
\end{align}
where the inequality is saturated for $\omega=0$ (for which we have $\nu=\cos\rho_{f}$). Then, we find
\begin{align}
J\leq \frac{N_{f}R^2}{\alpha'}  \frac{2}{\pi}
\int_{0}^{\rho_{f}} d\rho \frac{\cos\rho_{f}}{\sqrt{\sin^2\rho_{f} - \sin^2\rho}}\,.
\end{align}
This simply says that the folding point has the speed of light and so for a fixed string length $\rho_{f}$, the internal motion  is maximized when the string does not rotate inside $dS_3$. As depicted in Fig.~\ref{EJplot}, the right hand side is maximized in the short string limit $\rho_{f}\to 0$:
\begin{figure}[t] 
	\centering 
	\includegraphics[width=7.5cm]{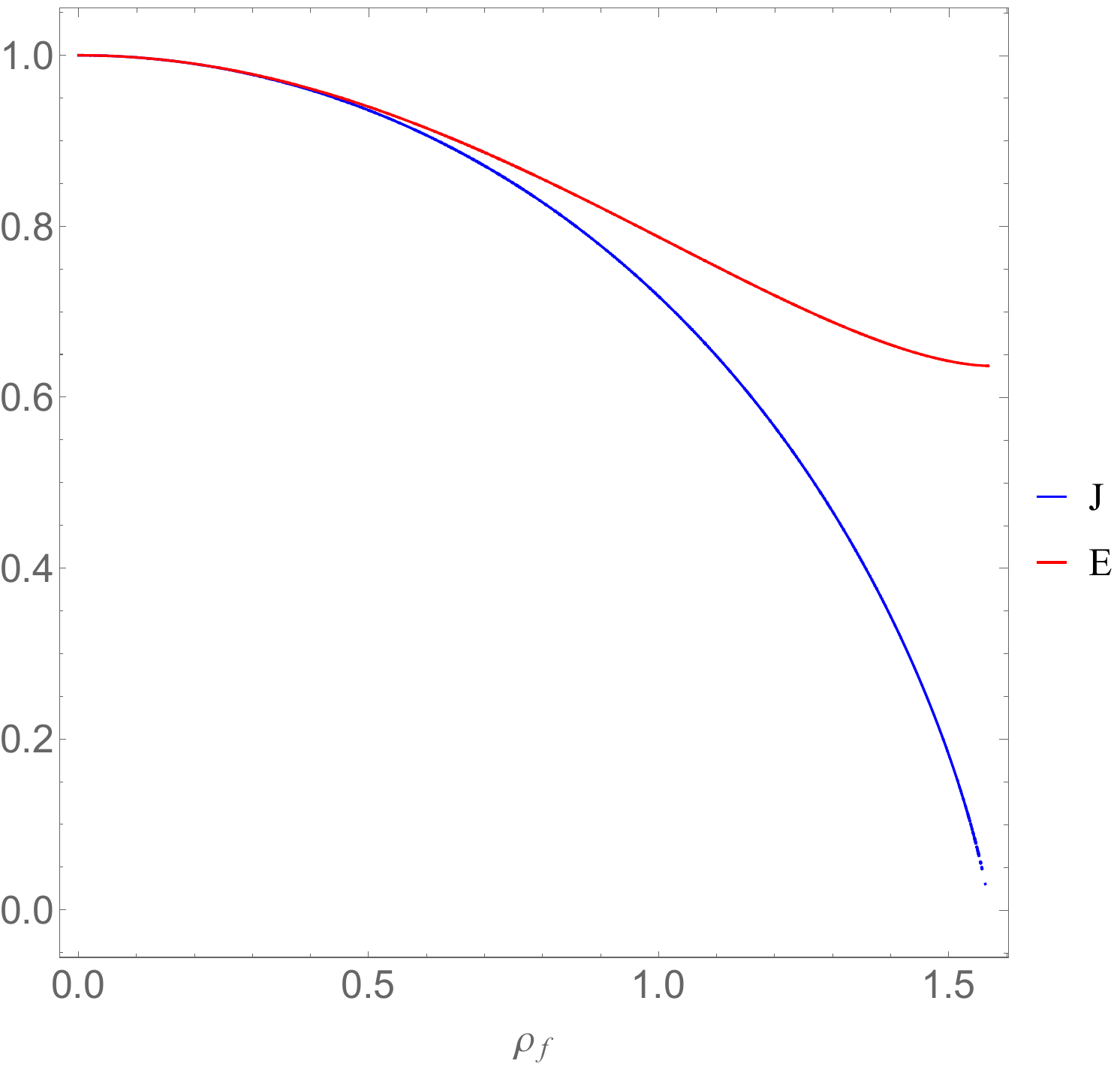}    
	\caption{The energy $E$ and the internal $U(1)$ charge $J$ as a function of $\rho_{f}$. They are plotted in the units of $N_{f} R/\alpha'$ and $N_{f} R^2/\alpha'$, respectively.} \label{EJplot}
\end{figure} 
\begin{align}
J\leq \frac{N_{f}R^2}{\alpha'}  \frac{2}{\pi}
\int_{0}^{\rho_{f}} d\rho \frac{\cos\rho_{f}}{\sqrt{\sin^2\rho_{f} - \sin^2\rho}}\leq \frac{N_{f}R^2}{\alpha'}\,.
\end{align}
Therefore, the $N_{f}$-folded string has the maximum internal charge $J=N_{f}R^2/\alpha'$ when $\omega=0$ and $\nu=\cos\rho_{f}\to 1$. Note that the energy $E$ and the spin $S$ in this limit are
\begin{align}
E=\frac{N_{f}R}{\alpha'}\,,
\quad
S=0\,,
\end{align}
which correspond to the upper boundary point of the shaded region in Fig.~\ref{folded_Regge}.

\subsection{Regge trajectories for fixed $J$}

\paragraph{Short strings.}

Next, let us take a closer look at the Regge trajectory profile for a fixed $J$. For this, we first consider the short string limit $\rho_{f}\ll 1$. In this regime, we have
\begin{align}
\rho_{f}\simeq \sqrt{\frac{1-\nu^2}{\omega^2+\nu^2}}\ll1\,,
\end{align}
so that the short string limit is realized for $\omega\gg1$, $\nu\simeq1$, or both (recall that causality requires $0\leq \nu^2\leq 1$). At the leading order in $\rho_{f}$, the charges~\eqref{E_folded_elliptic}-\eqref{J_folded_elliptic} are approximated~as
\begin{align}
	E & \simeq\frac{1}{\sqrt{1-\nu^2}}\times\frac{4N_{f}R}{2\pi \alpha'} \int_{0}^{\rho_{f}} d\rho \frac{1}{\sqrt{1 - (\rho/\rho_{f})^2 }}=\frac{1}{\sqrt{1-\nu^2}}\times\frac{N_{f}R}{\alpha'} \rho_{f}
    	~ , \label{short_fenergy}\\
	S &  \simeq\frac{\omega}{\sqrt{1-\nu^2}}\times\frac{4N_{f}R^2}{2\pi \alpha'} \int_{0}^{\rho_{f}} d\rho \frac{\rho^2}{\sqrt{1 - (\rho/\rho_{f})^2}}=\frac{1}{2} \sqrt{\frac{\omega^2}{\omega^2 + \nu^2}}\times\frac{N_{f}R^2}{\alpha'} \rho_{f}^2
~,\label{short_fspin}\\
	J &  \simeq\frac{\nu}{\sqrt{1-\nu^2}}\times\frac{4N_{f}R^2}{2\pi \alpha'} \int_{0}^{\rho_{f}} d\rho \frac{1}{\sqrt{1 - (\rho/\rho_{f})^2}}=\frac{\nu}{\sqrt{1-\nu^2}}\times\frac{N_{f}R^2}{\alpha'} \rho_{f}
	~. \label{short_fcharge}
\end{align}
Then, in the regime $
\omega\gg1$, which implies $J
\ll1$ in particular, 
we find the relation,
\begin{align}
    E^2\simeq \frac{J^2}{R^2}+\frac{2N_{f}}{\alpha'}S\,.
\end{align}
Recall that the short string limit is also achieved when $\omega = O(1)$ and $\nu\simeq 1$. In this regime, the internal charge~\eqref{short_fcharge} is not necessarily small because the prefactor $\frac{\nu}{\sqrt{1-\nu^2}}$ cancels out the suppression by the small $\rho_{f}$. Taking into account the next-to-leading order terms in Eqs.~\eqref{short_fenergy}, \eqref{short_fcharge} carefully, we find a more general energy-spin relation,
\begin{align}
    E^2\simeq \frac{J^2}{R^2}+{\sqrt{1-\left(\frac{\alpha^\prime}{N_{f}R^2} J\right)^2}}\,\frac{2N_{f}}{\alpha'}S\,, \label{flat_short_large}
\end{align}
which is applicable for an arbitrary value of $J$ as long as the string is short $\rho_{f}\ll1$. Note that the first term is the Kaluza-Klein mass associated to the internal motion, which explains how the Regge trajectory shifts upwards as $J$ increases.

\begin{figure}[t] 
	\centering 
	\includegraphics[width=8cm]{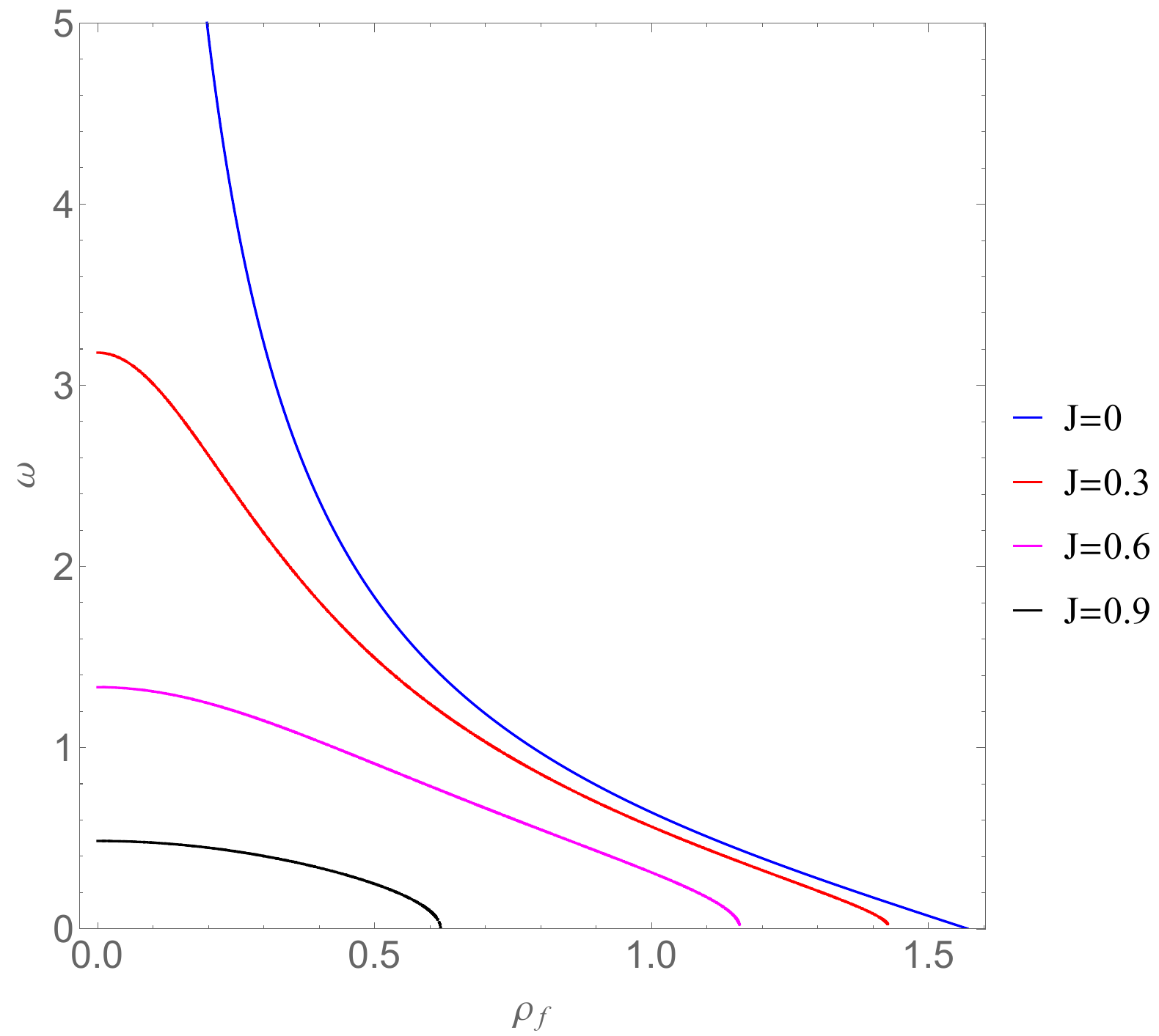}    
	\caption{The angular velocity $\omega$ as a function of $\rho_{f}$ ($J$ is in the unit of $N_{f} R^2/\alpha'$).} \label{rhofplot}
\end{figure}

\paragraph{Long strings.}

To discuss longer strings, it is convenient to rewrite Eq.~\eqref{J_folded_elliptic} as
\begin{align}
\omega^2=\frac
{\cot^2\rho_{f}\cdot\left[\frac{4N_{f}R^2}{2\pi \alpha'}  \mathcal{F}\left(\rho_{f} | \csc ^{2} \rho_{f}\right)\right]^2-J^2}
{\left[\frac{4N_{f}R^2}{2\pi \alpha'}  \mathcal{F}\left(\rho_{f} | \csc ^{2} \rho_{f}\right)\right]^2+J^2}\,,
\end{align}
where the right hand side monotonically decreases as $\rho_{f}$ increases (see Fig.~\ref{rhofplot}). It implies that  for a fixed $J$, there exists an upper bound on the angular velocity $\omega$: 
\begin{align}
0\leq \omega^2\leq \frac{\frac{N_{f}^2R^4}{\alpha'^2}-J^2}{J^2}\,,
\end{align}
where the upper bound is saturated in the short string limit $\rho_{f}\to 0$. Also, for a fixed $J$, the string has a maximum length when $\omega=0$, for which the conserved charges read
\begin{align}
\label{E_omega0}
	E & =\frac{1}{\sin\rho_{f}}\times\frac{4N_{f}R}{2\pi \alpha'} \left[\sin^2\rho_{f}\,\mathcal{E}\left(\rho_{f} | \csc ^{2} \rho_{f}\right)+\cos^2\rho_{f}\,\mathcal{F}\left(\rho_{f} | \csc ^{2} \rho_{f}\right)\right]
	~ , \\ 
	S &  =0
~,\\ 
\label{J_omega0}
	J &  =\cot\rho_{f}\times \frac{4N_{f}R^2}{2\pi \alpha'}  \mathcal{F}\left(\rho_{f} | \csc ^{2} \rho_{f}\right)\,.
\end{align} 
For a given $J$, the maximum length is determined by solving Eq.~\eqref{J_omega0}. Then, substituting it into Eq.~\eqref{E_omega0} gives the energy of the longest string. See also Fig.~\ref{EJplot}. This gives the upper endpoint of each Regge trajectory with a fixed $J$ depicted in Fig.~\ref{folded_Regge}.

\section{Spiky strings}\label{spikystrings}

Next, we study spiky strings (see Ref.~\cite{Kruczenski:2004wg} for spiky strings in AdS).
In this section we focus on the case without internal motion, so that our ansatz here is Eq.~\eqref{rigid_ansatz} with $\nu=\psi=0$, under which the equations of motion~\eqref{eom_rho}-\eqref{eom_phi} reduce to
\begin{align}
0&=-\partial_\sigma\left[\frac{\rho'(\cos^2\rho-\omega^2\sin^2\rho)}{\sqrt{(\cos^2\rho-\omega^2\sin^2\rho)\rho'^2+N^2\cos^2\rho\sin^2\rho}}\right]
\nonumber
\\
\label{spikyeom_rho}
&\quad
+\frac{1}{2}\frac{\sin2\rho[-(1+\omega^2)\rho'^2+N^2\cos2\rho]}{\sqrt{(\cos^2\rho-\omega^2\sin^2\rho)\rho'^2+N^2\cos^2\rho\sin^2\rho}}\,,
\\
0&={\partial_\sigma}\left[\frac{  \cos^2 \rho  \sin^2 \rho }{ \sqrt{(\cos^2\rho-\omega^2\sin^2\rho)\rho'^2+N^2\cos^2\rho\sin^2\rho}} \right]\,.\label{spikyeom_phi}
\end{align}
To follow the string dynamics, it is convenient to integrate Eq.~\eqref{spikyeom_phi} as
\begin{align}
\left|\rho'\right| &= \frac{{N}}{2} \frac{\sin 2\rho}{\sin 2 \rho_0} \sqrt{\frac{\sin^2 2 \rho - \sin^2 2\rho_0}{\cos^2\rho - \omega^2\sin^2\rho}}\,,
\end{align}
where the integration constant $\rho_0$ is chosen such that $\rho'=0$ for $\rho=\rho_0$. For later use,  we also define $\rho_1$ such that $\cot^2\rho_1=\omega^2$ and $0<\rho_1<\frac{\pi}{2}$. In this language, we have
\begin{align}
|\rho'|&={ \frac{N\sin\rho_1\sin 2\rho}{\sqrt{2}\sin 2 \rho_0}  \sqrt{\frac{\cos^2 2 \rho_0 - \cos^2 2\rho}{\cos2\rho-\cos 2\rho_1}}}
~. \label{spikyansatz}
\end{align}

\paragraph{Three shapes.}

Notice that $\rho'$ has to flip a sign somewhere in order for a closed string to form a loop, otherwise the string stretches forever. Such a sign flip may appear when $\rho'=0$ or $\rho'=\infty$. Eq.~\eqref{spikyansatz} shows that $\rho'=0$ is satisfied at $\rho=\rho_0,\frac{\pi}{2}-\rho_0$. At these points, the string smoothly turns back from inside to outside or vice versa. Without loss of generality, we assume $0<\rho_0<\frac{\pi}{4}$ in the following. On the other hand, $\rho'=\infty$ is satisfied at $\rho=\rho_1$, where the string turns back forming a spike. Based on the value of $\rho_1$ relative to $\rho_0$ and $\frac{\pi}{2} - \rho_0$, we may classify shapes of the string into the following three classes:
\begin{enumerate}

\item Outward spikes ($\rho_0<\rho_1< \frac{\pi}{2} - \rho_0$)

Recall that the inside of the square root in Eq.~\eqref{spikyansatz} has to be positive for $\rho$ to be real. Therefore, the reality condition implies that for this parameter set, the string may stretch only inside the region $\rho_0\leq\rho\leq\rho_1$. This means that the outer turning points are spiky, and the inner ones are smooth.
We call such strings outward spike solutions. See Fig.~\ref{SpikyStringdS}.

\item Rounded spikes ($\rho_0<\frac{\pi}{2} - \rho_0<\rho_1$)

Similarly, for $\frac{\pi}{2}-\rho_0<\rho_1<\frac{\pi}{2}$, the reality condition implies that the string may stretch only inside the region $\rho_0\leq\rho\leq\frac{\pi}{2} - \rho_0$. In contrast to the case of outward spikes, strings in this class have no spikes and all the turning points are smooth.  We call such strings rounded spike solutions. See Fig.~\ref{circularround}.  Note that these strings are specific to de Sitter space and there are no counterpart in flat space and AdS.

\item Inward spikes ($\rho_1<\rho_0< \frac{\pi}{2} - \rho_0$)

Finally, for $\rho_1<\rho_0$, the string may stretch only inside the region $\rho_1\leq\rho\leq\rho_0$.  This means that the outer turning points are smooth, and the inner ones are spiky. We call such strings inward spike solutions. See Fig.~\ref{shape_inward}.

\end{enumerate}

\begin{figure}[t] 
	\centering 
	\includegraphics[width= 15 cm]{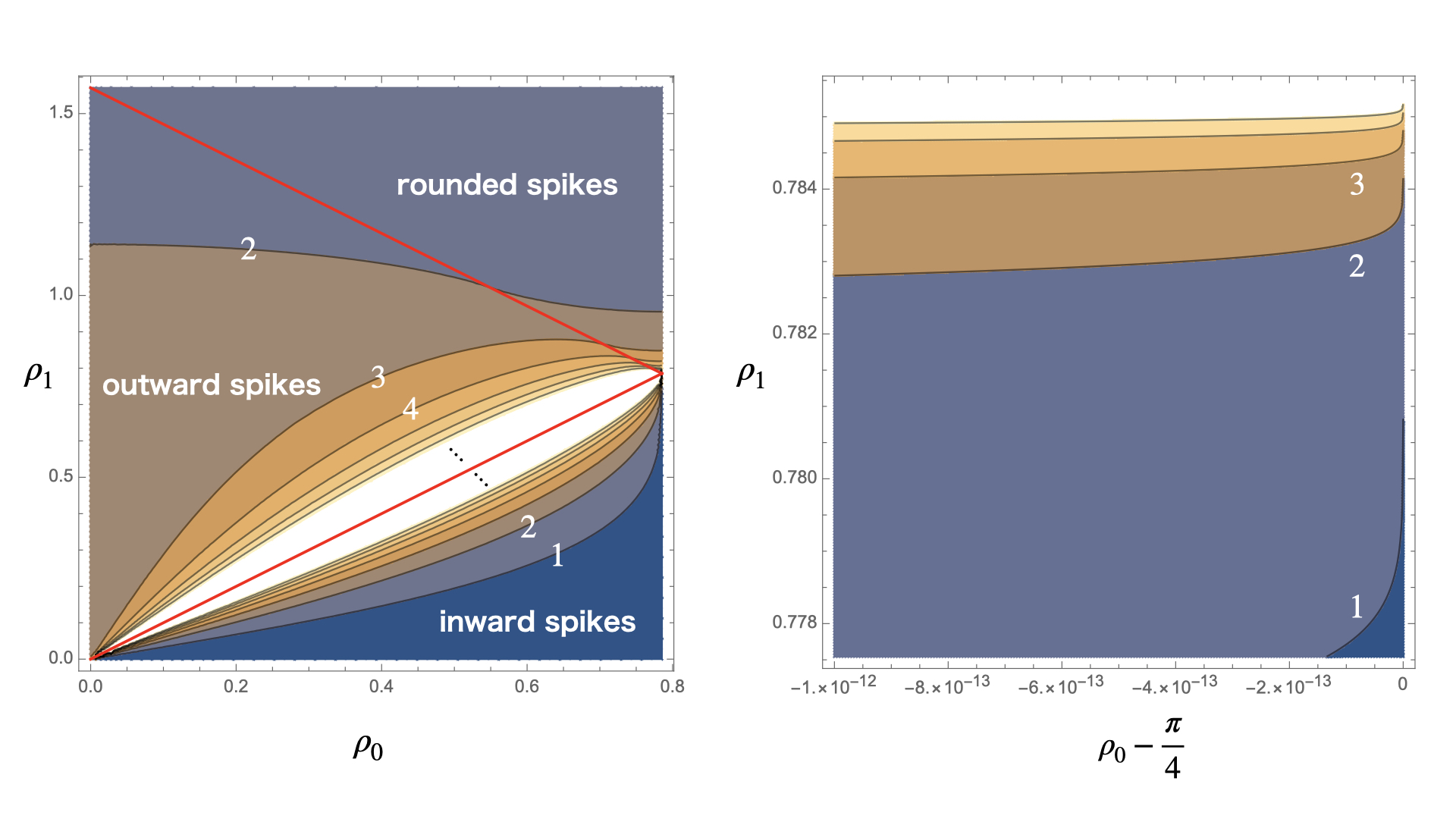}    	\caption{
Contour plot of $2\pi/\Delta\phi$ as a function of $\rho_0$ and $\rho_1$: An integer on each contour represents the value of $2\pi/\Delta\phi$ for given $\rho_0$ and $\rho_1$, which has to be $n/N$ for $n$-spike strings with the winding number $N$. The two red lines, $\rho_1=\rho_0$ and $\rho_1=\frac{\pi}{2}-\rho_0$, separate the $\rho_0$ -- $\rho_1$ plane into three regions which accommodate outward spikes, rounded spikes, and inward spikes, respectively. We find in particular that the string shape has a smooth transition from outward spikes to rounded spikes, as $\rho_0$ increases from $0$ to $\pi/4$ for fixed $n$ and $N$. Another important observation is that for inward spikes, $\rho_1 < \frac{\pi}{4}$ at $\rho_0 = \frac{\pi}{4}$ for a finite $2\pi/\Delta \phi$ (see the right zoom-in figure around $(\rho_0,\rho_1) = (\frac{\pi}{4},\frac{\pi}{4})$). 
} \label{Fig:rho0rho1}
\end{figure}

\paragraph{Periodicity conditions.}

The above argument is useful enough to classify local shapes of the string. On the other hand, the full string is made of multiple segments between the spikes. In order for a closed string to form a loop, the angle $\Delta\phi$ of each segment has to be quantized appropriately. For our ansatz, an explicit form of $\Delta\phi$ is given by
\begin{align}\nonumber
\Delta\phi&={2N}\int_{\rho_{\min}}^{\rho_{\max}}  \frac{d\rho}{\rho'}\\*
&={2\int_{\rho_{\min}}^{\rho_{\max}} \,d\rho\,
 \frac{\sqrt{2}\sin 2 \rho_0}{\sin\rho_1\sin 2\rho}  \sqrt{\frac{\cos2\rho-\cos 2\rho_1}{\cos^2 2 \rho_0 - \cos^2 2\rho}}} \label{angle_neighbor}\,,
\end{align}
where $\rho_{\min}$ and $\rho_{\max}$ are the minimum and the maximum values of $\rho$. 
More explicitly, $(\rho_{\rm mi{n}},\rho_{\rm max})=(\rho_0,\rho_1),(\rho_0,\frac{\pi}{2}-\rho_0),(\rho_1,\rho_0)$ for outward spikes, rounded spikes, and inward spikes, respectively.
Then, the global consistency requires that
\begin{align}
\Delta\phi=\frac{2\pi N}{n}\,, \label{perio_spiky}
\end{align}
where $n$ is a positive integer characterizing the number of spikes. This determines the value of $\rho_1$ for given $\rho_0$, $n$, and $N$. See also Fig.~\ref{Fig:rho0rho1} for a plot of $2\pi/\Delta\phi$ as a function of $\rho_0$ and $\rho_1$, which shows a smooth transition from outward spikes to rounded spikes for fixed $n$ and $N$.

\paragraph{Energy and spin. }

For later convenience, we provide the energy and the spin \eqref{E_master}-\eqref{S_master} for the present class of solutions by using Eq.~\eqref{spikyansatz} as
\begin{align}
E  & = \frac{\omega S}{R} + \frac{R}{2\pi\alpha' }(2n)\int_{\rho_{\min}}^{\rho_{\max}} d\rho  \sin 2\rho \frac{\sqrt{\cos^2 \rho-\omega^2 \sin^2 \rho}}{\sqrt{\sin^2 2 \rho - \sin^2 2 \rho_0}}\,,\label{spiky_def_energy}\\
S &  =  \frac{R^2}{2\pi\alpha'} \times \frac{1}{2} (2n) \int_{\rho_{\min}}^{\rho_{\max}}  d\rho  \frac{\omega\sin\rho}{\cos \rho }   \frac{\sqrt{\sin^2 2 \rho - \sin^2 2\rho_0}}{\sqrt{\cos^2\rho - \omega^2\sin^2\rho}}\,,  
\end{align}
where we used Eq.~\eqref{worldsh_Ham} to derive Eq.~\eqref{spiky_def_energy}. In the rest of the section, we study the three types of string solutions in more details.

\subsection{Outward spike solutions}

\begin{figure}[t] 
	\centering 
	\includegraphics[width=6.35cm]{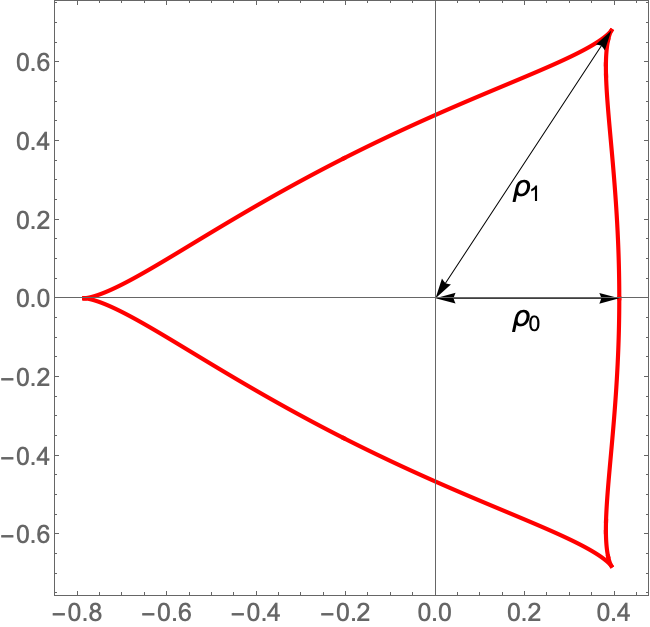}    \quad  \quad  \quad 
	\includegraphics[width=6.35cm]{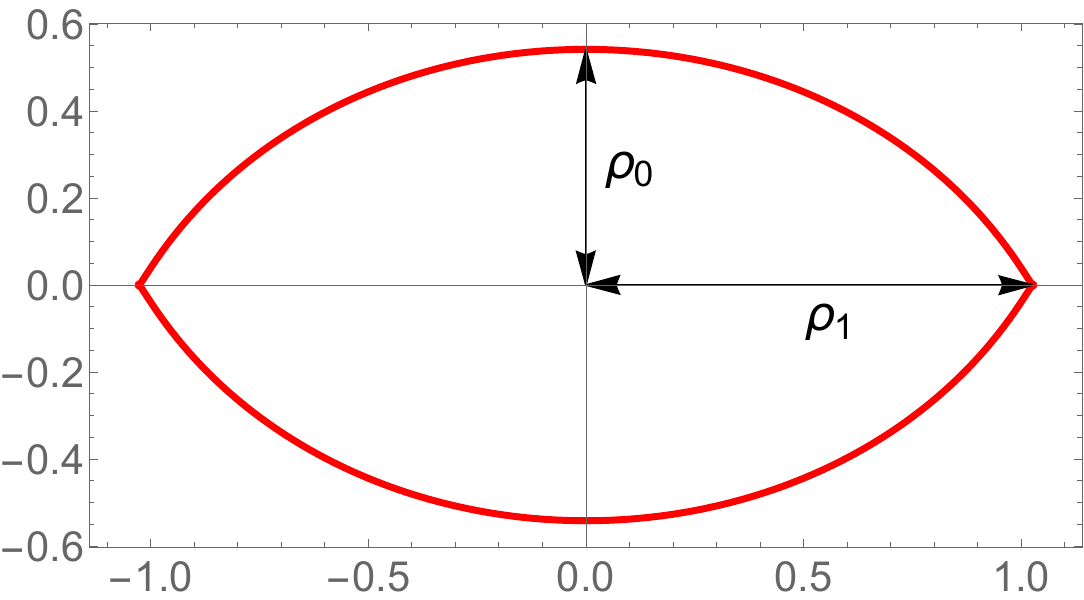}   
	\caption{Typical shapes of outward spike solutions.  The left panel shows the solution with three outward spikes and one winding for $\rho_0 \simeq 0.41$ and $\rho_1 = \frac{\pi}{4}$. The right panel shows the solution with two outward spikes and one winding for $\rho_0 \simeq 0.54$ and $\rho_1 \simeq 1.03$. The latter type of solutions are specific to de Sitter space.} \label{SpikyStringdS}
\end{figure}
\begin{figure}[t] 
	\centering 
	\includegraphics[width=6.7cm]{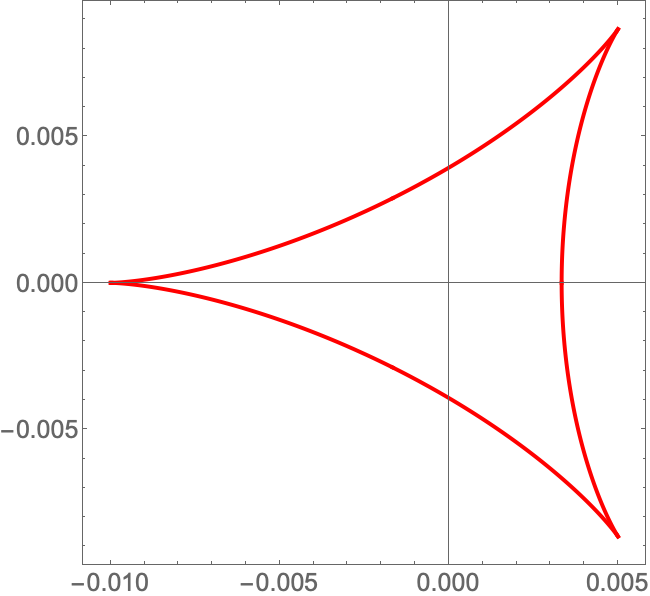}    \quad  \quad  \quad 
	\includegraphics[width=6.35cm]{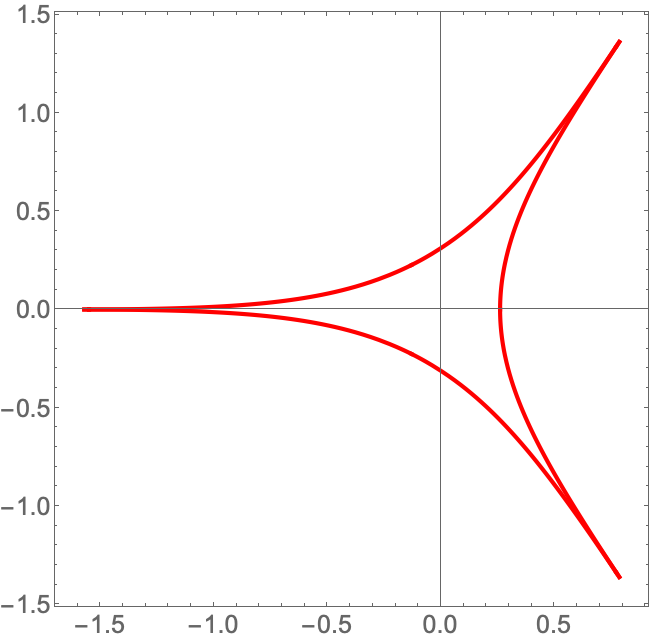}  
	\caption{Typical shapes of spiky strings with three outward spikes in flat space (the left panel) and AdS (the right panel).
	 } \label{comparisonadsds}
\end{figure}

\begin{figure}[t] 
	\centering 
	\includegraphics[width=5.5cm]{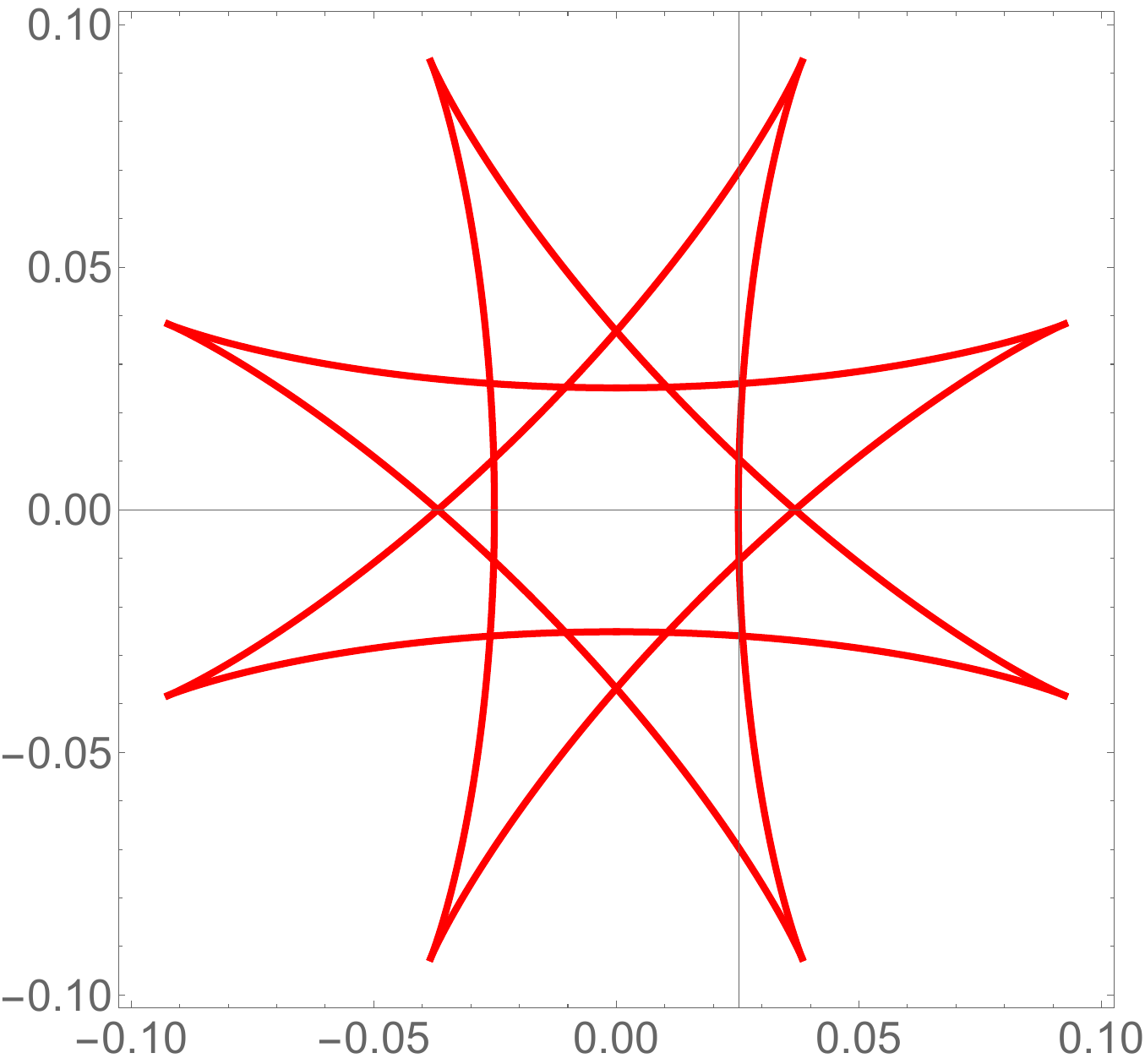} \quad \quad
	\includegraphics[width=7.5cm]{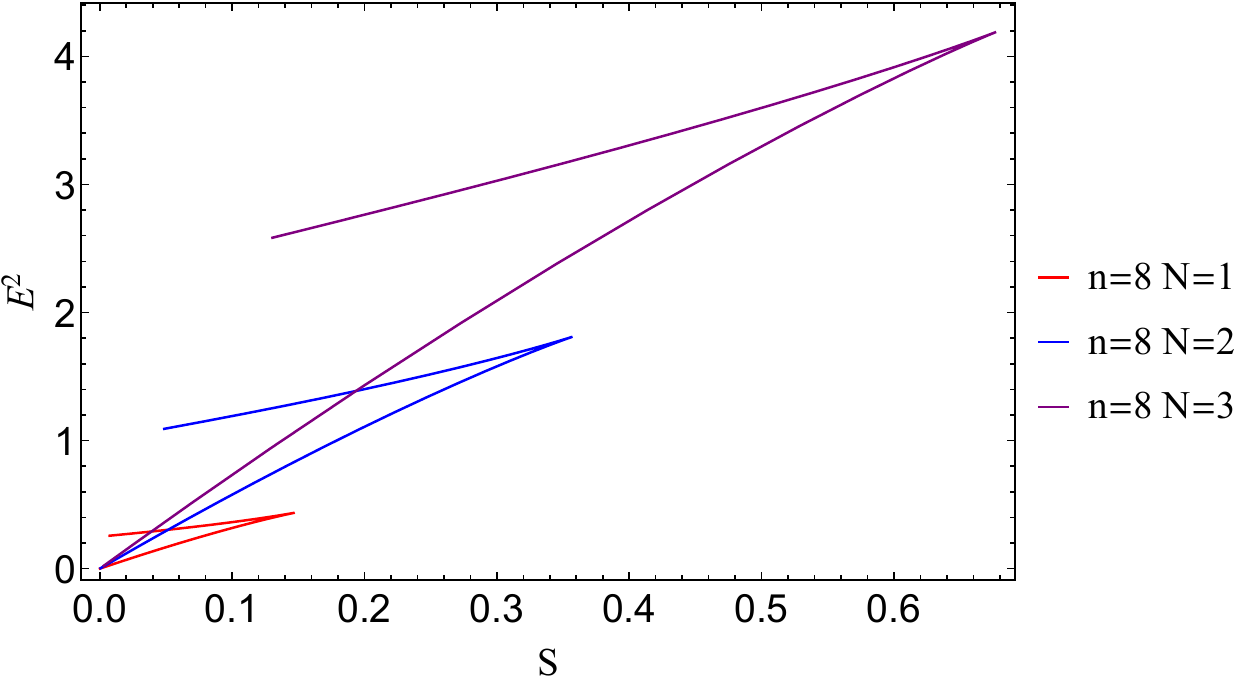}     
	\caption{The left panel shows the shape of spiky string  with $8$ outward spikes and $3$ windings for $\rho_0 \simeq 0.025$ and $\rho_1 = 0.1$. The right panel shows the Regge trajectories of different windings. The spin $S$ and the energy $E$ are plotted in the units of $R^2/\alpha'$ and $R/\alpha'$, respectively.} \label{NN3n8}
\end{figure}

We begin with outward spike solutions ($\rho_0 < \rho_1< \frac{\pi}{2} - \rho_0$), whose typical shapes are given in Fig.~\ref{SpikyStringdS}. See also the left panel of Fig.~\ref{NN3n8} for strings with more windings. To identify the shapes, first we  derive a relation between $\rho_0$ and $\rho_1$.  If the number of spikes $n$ and the winding number $N$ are specified, we may derive the relation from  the periodicity condition \eqref{angle_neighbor}-\eqref{perio_spiky} as
\begin{align}
\frac{2\pi N }{n}
={2\int_{\rho_0}^{\rho_1}  d\rho
 \frac{\sqrt{2}\sin 2 \rho_0}{\sin\rho_1\sin 2\rho}  \sqrt{\frac{\cos2\rho-\cos 2\rho_1}{\cos^2 2 \rho_0 - \cos^2 2\rho}}\,.} \label{perio_out2}
\end{align}
Now, we are left with one parameter $\rho_0$, which  characterizes the size of the string. If we further specify $\rho_0$, we may identify the shape of the string simply by integrating
\begin{align}
\frac{d\phi}{d\rho}=\pm\frac{\sqrt{2}\sin 2 \rho_0}{\sin\rho_1\sin 2\rho}  \sqrt{\frac{\cos2\rho-\cos 2\rho_1}{\cos^2 2 \rho_0 - \cos^2 2\rho}} \,. \label{phi_rho}
\end{align}
For example, the plots in Fig.~\ref{SpikyStringdS} are obtained in this way. It is also instructive to compare the shapes there with those in flat space and AdS. See Fig.~\ref{comparisonadsds}. We find that in de Sitter space, the inner turning points shift outward compared to the flat space case due to de Sitter acceleration whereas in AdS, the inner turning points shift inward due to AdS deceleration. In particular, the $n=2N$ case reflects this effect most clearly: As depicted in the right figure of Fig.~\ref{SpikyStringdS}, de Sitter space accommodates spiky strings which can be thought of as a fatter version of the folded strings. Both in flat space and AdS, such a spiky string is not stable because the string tension always overcomes the centrifugal force, so that it collapses to the folded string. In sharp contrast, de Sitter acceleration helps the spiky string to maintain the shape without collapsing into a folded string.

\begin{figure}[t] 
	\centering 
	\includegraphics[width=8cm]{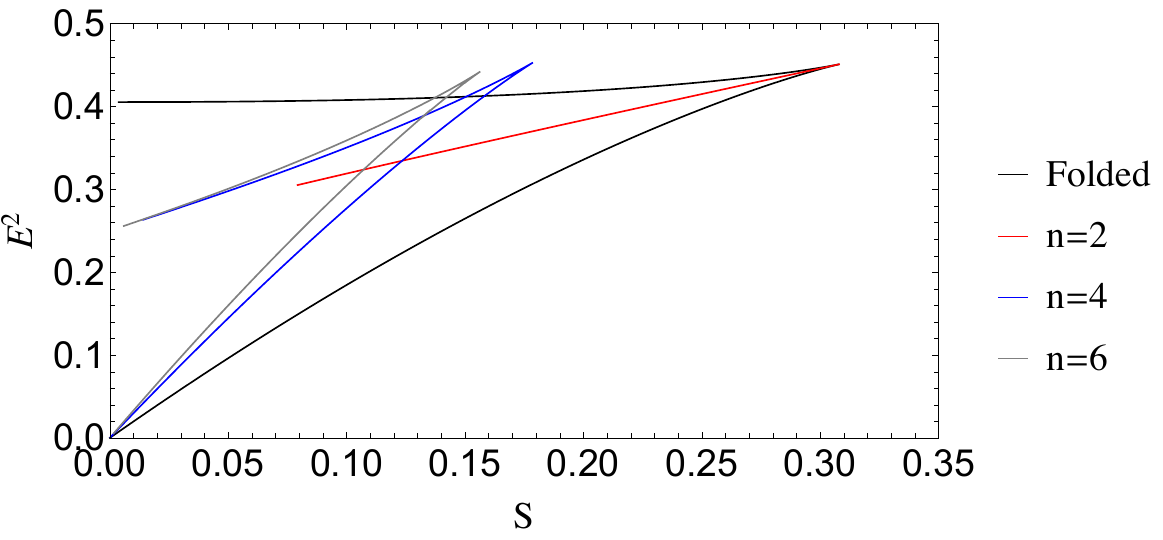}    \\ 
	\includegraphics[width=8cm]{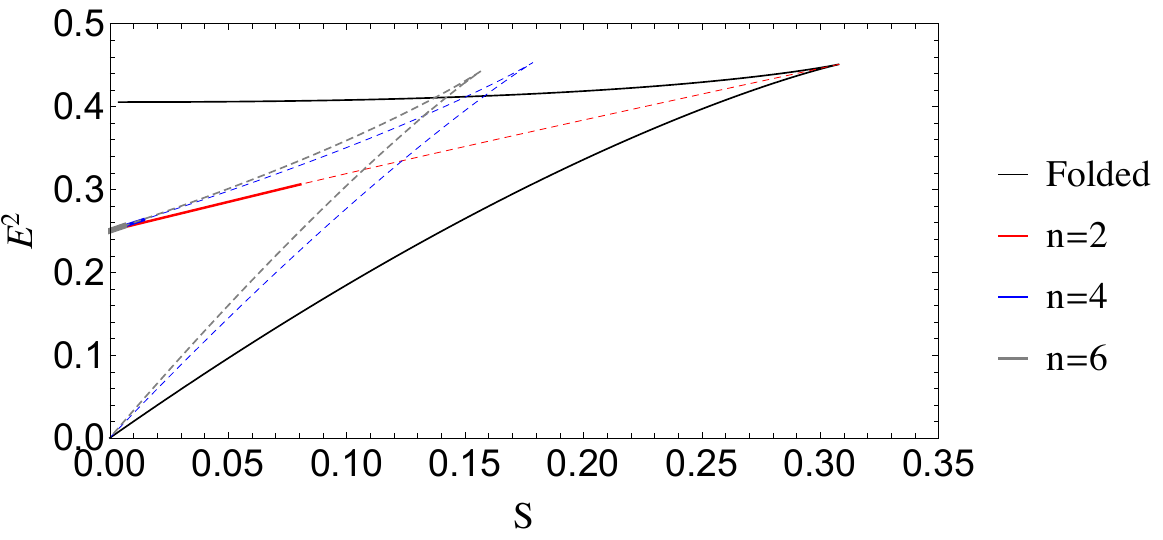}  
	\caption{Regge trajectories of outward spike solutions (the upper panel) and rounded spike solutions (the lower panel) for the winding number $N=1$. The spin $S$ and the energy $E$ are plotted in units of $R^2/\alpha'$ and $R/\alpha'$, respectively, as before. For comparison, we also illustrate the Regge trajectory of one-folded strings. The dotted curves in the lower panel are the Regge trajectories for outward spikes, which are smoothly connected with those for rounded spikes. } \label{SpikyStringOutRegge}
\end{figure} 

\paragraph{Regge trajectories.}

Using the $\rho_0$ -- $\rho_1$ relation~\eqref{perio_out2}, we can calculate the energy $E$ and the spin $S$ as a function of $\rho_0$, which defines Regge trajectories. See Fig.~\ref{SpikyStringOutRegge} for those of winding number $N=1$ solutions. First, we find that each trajectory has an approximately linear form up to the maximum spin point and then it turns back, similarly to the folded string case. 
In particular, the spin at the turning point is smaller than that of  the folded string. As a result, the spectrum satisfies  the Higuchi bound. We also find that the tilt in the linear region is steeper for strings with a larger number of spikes. Second, the upper endpoint of the Regge trajectory does not touch the vertical axis $S=0$ in contrast to the folded string case. In the next subsection, we show that the trajectory is smoothly connected to that of rounded spike solutions, which touches the vertical axis $S=0$. Third, spiky strings with a fixed winding number $N$ scan a finite region of the energy-spin plane. Therefore, to obtain solutions with a larger spin, we need to increase the winding number $N$. See Fig.~\ref{NN3n8}.

\medskip
Besides, another remark is needed on the Regge trajectory of $n=2N$ solutions. See the red curve in the upper panel of Fig.~\ref{SpikyStringOutRegge} for $n=2$ and $N=1$. As we mentioned, the $n=2N$ solutions can be thought of as a fatter version of folded strings, which are supported by de Sitter acceleration. Then, one may expect that such solutions collapse into folded strings when the string is small and so the support of de Sitter acceleration is not enough. Indeed, we find that the Regge trajectory of $n=2N$ outward spikes branches from the turning point of the folded string trajectory.

\paragraph{Short strings.}

To provide more quantitative discussion, let us study the short string regime of outward spike solutions:
\begin{align}
\rho_0\,,\; \rho_1 \ll \frac{\pi}{2}\,.
\end{align}
For such short strings, the $\rho_0$ -- $\rho_1$ relation~\eqref{perio_out2} is approximated as
\begin{align}
\frac{2\pi N }{n}
\simeq 
2\, \frac{\rho_0}{\rho_1} \int_{\rho_{0}}^{\rho_{1}} \frac{d \rho}{\rho}\frac{\sqrt{\rho_1^2 - \rho^2}}{\sqrt{\rho^2- \rho_0^2}} = \frac{\rho_1 - \rho_0}{\rho_1}\pi \quad
\leftrightarrow\quad
\left( 1 - \frac{2N}{n} \right) \rho_1 \simeq \rho_0 \,. \label{outward_short01}
\end{align}
This shows that $\rho_0=0$ for $n=2N$ at least under the short string approximation, which is consistent with the fact that the $n=2N$ solutions are extrapolated to folded strings as they become smaller.
Also, the energy and spin are approximated as
\begin{align}
E&\simeq \frac{R}{2\pi\alpha' } \times 2n \int_{\rho_{0}}^{\rho_{1}} d\rho \frac{\rho (\rho_1^2-\rho_0^2)}{\rho_1\sqrt{\rho^2-\rho_0^2}\sqrt{\rho_1^2-\rho^2}} =\frac{nR}{2\alpha^\prime} \frac{\rho_1^2 - \rho_0^2}{\rho_1} =2N\left( 1 -\frac{N}{n}\right) \frac{R}{\alpha^\prime} \rho_1 \,,
\\*
S &\simeq \frac{R^2}{2\pi \alpha' } \times 2n \int_{\rho_{0}}^{\rho_{1}} d\rho \ \rho \frac{\sqrt{\rho^2 - \rho_0^2}}{\sqrt{\rho_1^2 - \rho^2}}  = \frac{n R^2}{4\alpha'} (\rho_1^2-\rho_0^2) = N \left( 1 -\frac{N}{n} \right) \frac{R^2}{\alpha^\prime} \rho_1^2\,,
\end{align}
from which the energy-spin relation reads
\begin{align}
E^2 \simeq \frac{4}{\alpha^\prime} N \left( 1 - \frac{N}{n} \right)S \,. \label{short_regge_out}
\end{align}
This correctly reproduces the linear Regge trajectory in flat space. We find that the tilt of the Regge trajectory is steeper for a larger number of spikes.  In particular, in the limit of infinitely many spikes (for $N$ fixed), the tilt approaches to $\frac{4}{\alpha^\prime} N$. We will find in Sec.~\ref{Subsec:inward} that steeper Regge trajectories are realized by inward spike solutions.

\paragraph{Long strings.}

Finally, let us take a closer look at the long string regime. First, the condition $\rho_0 < \rho_1 < \frac{\pi}{2}- \rho_0$ of outward spikes implies that $\rho_0$ cannot be larger than ${\pi}/{4}$.  Also, as $\rho_0$ approaches to $\pi/4$, $\rho_1$ approaches to $\pi/4$ and so the spin decreases essentially because the closed string becomes nearly circular and the change of the worldsheet profile by rotation becomes smaller. To interpolate the short string regime, where the spin increases, and the long string regime, where the spin decreases, the Regge trajectory needs to have the maximum spin. 

\medskip
More quantitatively, the maximum value of $\rho_0$ depends on the number of spikes $n$ and the winding number $N$. As we mentioned earlier, we obtain the $\rho_0$ -- $\rho_1$ relation~\eqref{perio_out2} depicted in Fig.~\ref{Fig:rho0rho1}, once $n$ and $N$ are specified. As we increase $\rho_0$ for given $n$ and $N$, each curve on the $\rho_0$ -- $\rho_1$ plane enters the rounded spike regime at some critical value and so there exists a smooth transition from outward spikes to rounded spikes. For example, the critical value for $n=4$ and $N=1$ reads $\rho_0 \simeq 0.75$, which corresponds to the upper endpoint of the Regge trajectory (see Fig.~\ref{SpikyStringOutRegge}). Beyond the critical value, the Regge trajectory describes rounded spike solutions, which we study in the next subsection.

\subsection{Rounded spike solutions}

\begin{figure}[t] 
	\centering 
	\includegraphics[width=7cm]{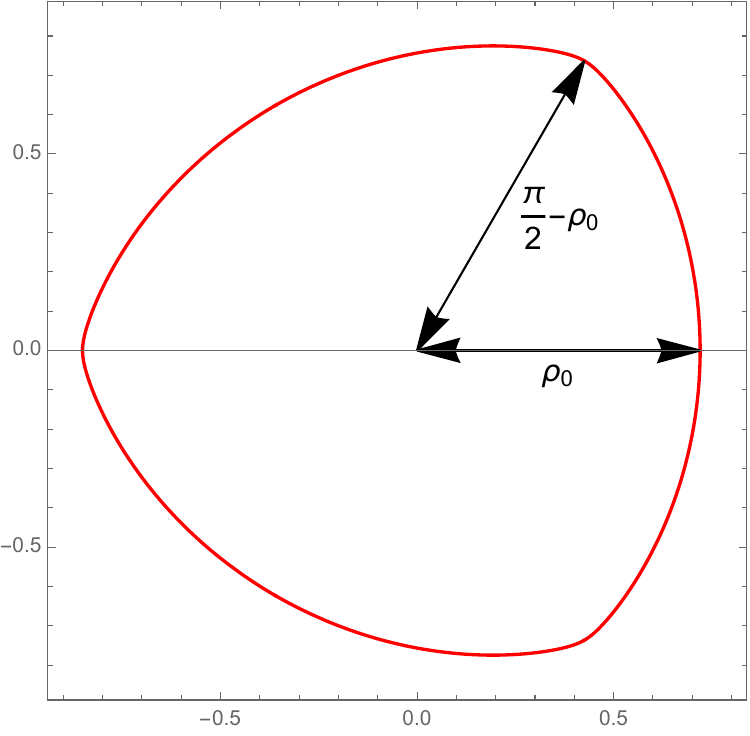}    
	\caption{Rounded spike solution for $\rho_0 = 0.72$ and $\rho_1 \simeq 0.86$. We call the turning points defined by $\rho = \pi/2 - \rho_0$ rounded spikes.} \label{circularround}
\end{figure} 

Next, we discuss rounded spike solutions ($\rho_0 < \frac{\pi}{2} - \rho_0 < \rho_1$). See Fig.~\ref{circularround} for a typical shape of the string, which is regular everywhere. As we have just mentioned, this class of solutions are smooth continuation of outward spike solutions. Then, we may interpret that outward spikes for $\rho_1<\frac{\pi}{2} - \rho_0$ are rounded when $\rho_0$ crosses the critical value defined by $\rho_1=\frac{\pi}{2} - \rho_0$ (for given $n$ and $N$). Based on this interpretation, we call solutions with $\rho_1>\frac{\pi}{2} - \rho_0$  rounded spike solutions.

\medskip
The procedure to identify the shape is parallel to the case of outward spikes. First, we specify  the number of spikes $n$ and the winding number $N$, and derive a relation between $\rho_0$ and $\rho_1$ from the  periodicity condition,
\begin{align}
\frac{2\pi N }{n}
={2\int_{\rho_0}^{\pi/2 - \rho_0}   d \rho\,
 \frac{\sqrt{2}\sin 2 \rho_0}{\sin\rho_1\sin 2\rho}  \sqrt{\frac{\cos2\rho-\cos 2\rho_1}{\cos^2 2 \rho_0 - \cos^2 2\rho}}\,.} \label{perio_rounded2}
\end{align}
Then, integrating Eq.~\eqref{phi_rho}, we may identify the shape of the string for each $\rho_0$. Notice that this type of solutions do not exist for small $\rho_0$. See Fig.~\ref{Fig:rho0rho1}. For example, the allowed parameter range of $\rho_0$ for $n=4$ and $N=1$ reads  $0.75 \lesssim \rho_0 < \frac{\pi}{4}$.

\paragraph{Regge trajectories.}

Varying the value of $\rho_0$, we may draw the Regge trajectories as depicted in the right panel of  Fig.~\ref{SpikyStringOutRegge}. There, for comparison, we also illustrate the Regge trajectories of outward spike solutions by the dotted lines. Since rounded spikes exhibit a smooth transition to outward spikes, the Regge trajectories are connected with those of outward spikes. We also find that each Regge trajectory touches the vertical axis $S=0$, similarly to the folded string. However, as we discuss in the next paragraph, the mechanism how the spin vanishes is different from the folded string.

\paragraph{Circular string limit.} \label{circular}

To see how the spin vanishes, let us consider the limit $\rho_0\rightarrow \frac{\pi}{4}$. 
Recalling that $\rho_0\leq\rho(\sigma)\leq\frac{\pi}{2}-\rho_0$, we find that in this limit, the solution is reduced  to
\begin{align} 
\rho(\sigma)=\rho_0=\frac{\pi}{4}\,\,({\rm constant})\,,
\end{align}
which is nothing but the static circular string studied in Ref.~\cite{deVega:1993rm}. As discussed there, such a static circular string solution exists in de Sitter space because the string tension and the de Sitter acceleration balance and cancel each other out. Note that the equations of motion~\eqref{spikyeom_rho}-\eqref{spikyeom_phi} are satisfied for an arbitrary value of $\omega$, since rotations do not change the worldsheet profile and so they are gauge degrees of freedom. The conserved charges~\eqref{E_master}-\eqref{S_master} for these circular strings read
\begin{align}
E=\frac{NR}{2\alpha'}\,,
\quad
S=0\,.
\end{align}
In particular, the string has no spin for an arbitrary $\omega$ because the circular string has no structures generating nonzero angular momenta. 

\subsection{Inward spike solutions}
\label{Subsec:inward}

\begin{figure}[t] 
	\centering 
        \includegraphics[width=6.35cm]{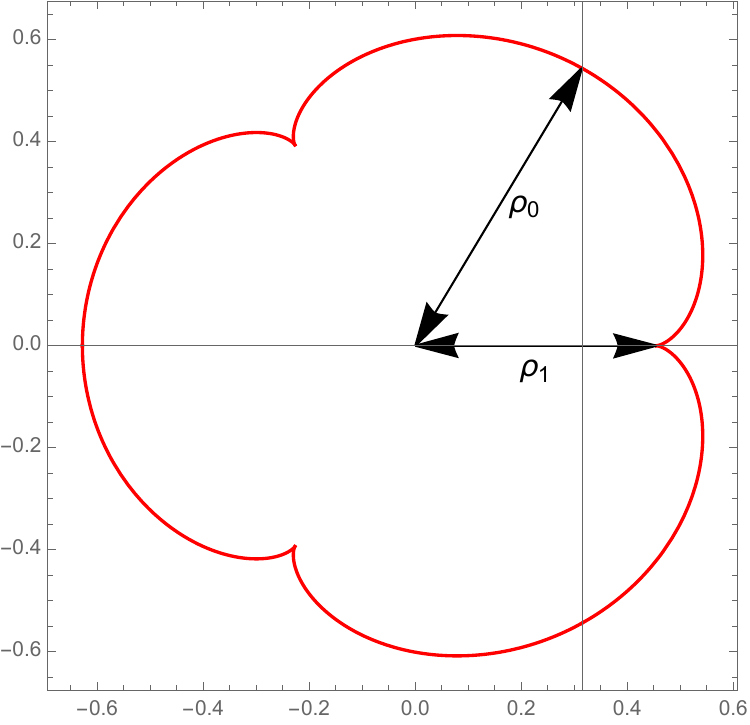}      \quad  \quad  \quad 
	\caption{Solution with three inward spikes and one winding for $\rho_0 = \pi/5$ and $\rho_{1}\simeq 0.46$.} \label{shape_inward}
\end{figure}

Finally, we discuss inward spike solutions ($\rho_1<\rho_0< \frac{\pi}{2} - \rho_0$), whose typical shape is illustrated in Fig.~\ref{shape_inward}. The procedure to identify the shape is parallel to the case of outward and rounded spikes. First, we specify  the number of spikes $n$ and the winding number $N$ and derive a relation between $\rho_0$ and $\rho_1$ from the  periodicity condition,
\begin{align}
\frac{2\pi N }{n}
={2\int_{\rho_1}^{\rho_0}   d \rho\,
 \frac{\sqrt{2}\sin 2 \rho_0}{\sin\rho_1\sin 2\rho}  \sqrt{\frac{\cos2\rho-\cos 2\rho_1}{\cos^2 2 \rho_0 - \cos^2 2\rho}}\,.} \label{perio_in2}
\end{align}
Then, by integrating Eq.~\eqref{phi_rho} for a specific value of $\rho_0$, we may identify the shape.

\begin{figure}[t] 
	\centering 
	\includegraphics[width=8cm]{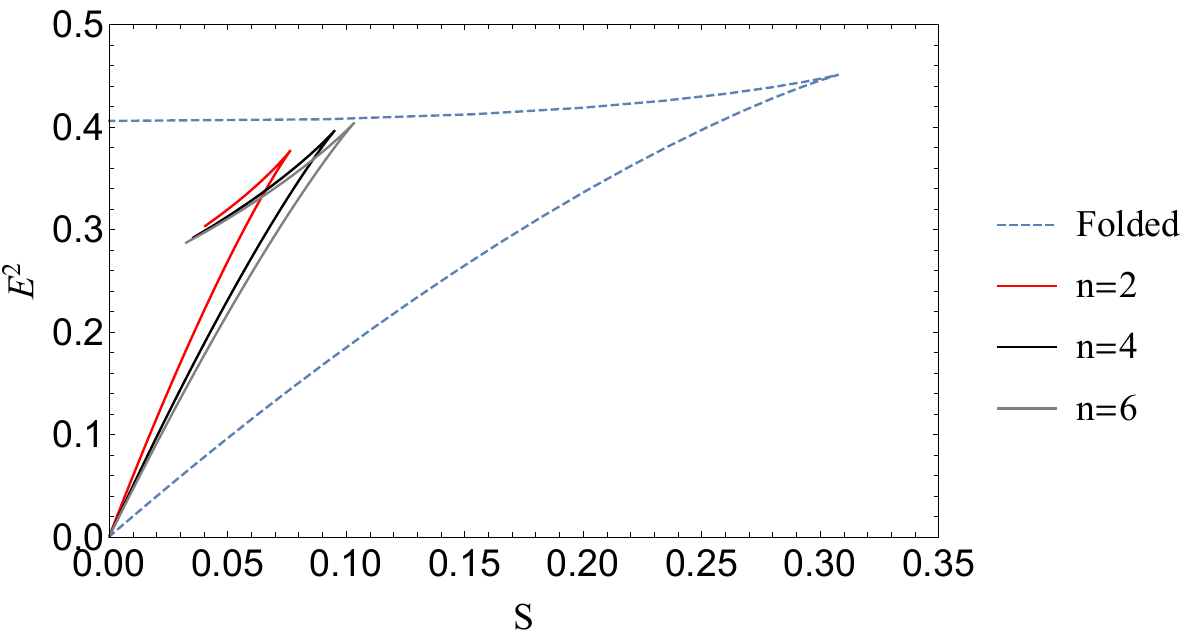}  
	\caption{Regge trajectories of inward spike solutions for $N = 1$. The spin $S$ and the energy $E$ are plotted in the units of $R^2/\alpha'$ and $R/\alpha'$, respectively. For comparison, we illustrate the Regge trajectory of the one-folded string by the dotted blue curve.} \label{SpikyStringdSRegge}
\end{figure}

\paragraph{Regge trajectories.}
The Regge trajectories are illustrated in Fig.~\ref{SpikyStringdSRegge}. Similarly to the previous cases, each Regge trajectory has the maximum energy and spin, 
which is helpful for the spectrum to satisfy the Higuchi bound. 
In contrast to outward spikes, the tilt in the short string regime decreases as the number of spikes increases. However, the tilt is always steeper than those of outward spike solutions and folded strings, as we discuss in the next paragraph in more details.
Note that the Regge trajectory does not touch the vertical axis $S=0$. As far as we know, there are no solutions at least within our ansatz that extrapolate the trajectory to $S=0$, in contrast to the outward spike case.

\paragraph{Short strings.}

Then, let us take a closer look at the short string regime:
\begin{align}
\rho_0\,,\; \rho_1 \ll \frac{\pi}{2}\,.
\end{align}
For such strings, the $\rho_0\,$ -- $\,\rho_1$ relation~\eqref{perio_in2} is approximated as
\begin{align}
\frac{2\pi N }{n}
\simeq 
2\, \frac{\rho_0}{\rho_1} \int_{\rho_{1}}^{\rho_{0}} \frac{d \rho}{\rho}\frac{\sqrt{\rho_1^2 - \rho^2}}{\sqrt{\rho^2- \rho_0^2}} = \frac{\rho_0 - \rho_1}{\rho_1}\pi \quad
\leftrightarrow\quad
\left( 1 + \frac{2N}{n} \right) \rho_1 \simeq \rho_0 \,, \label{inward_short01}
\end{align}
which implies that spiky strings can have an arbitrary number of inward spikes $n$ and an arbitrary winding number $N$ (recall that inward spike solutions in the short string regime have a condition $n>2N$). Also, the energy and the spin are approximated as
\begin{align}
	E&\simeq \frac{R}{2\pi\alpha' } \times 2n \int_{\rho_{1}}^{\rho_{0}} d\rho \frac{\rho (\rho_1^2-\rho_0^2)}{\rho_1\sqrt{\rho^2-\rho_0^2}\sqrt{\rho_1^2-\rho^2}} =\frac{nR}{2\alpha^\prime} \frac{\rho_0^2 - \rho_1^2}{\rho_1} =2N\left( 1  + \frac{N}{n}\right) \frac{R}{\alpha^\prime} \rho_1 \,,
\\
	S &\simeq \frac{R^2}{2\pi \alpha' } \times 2n \int_{\rho_{1}}^{\rho_{0}} d\rho \ \rho \frac{\sqrt{\rho^2 - \rho_0^2}}{\sqrt{\rho_1^2 - \rho^2}}  = \frac{n R^2}{4\alpha'} (\rho_0^2-\rho_1^2) 
= N \left( 1 + \frac{N}{n} \right) \frac{R^2}{\alpha^\prime} \rho_1^2\,,
\end{align}
which reproduce the linear Regge trajectories in flat space,
\begin{align}
E^2 \simeq \frac{4}{\alpha^\prime} N \left( 1 +  \frac{N}{n} \right)S \,. \label{short_regge_in}
\end{align}
We find that the tilt of the Regge trajectory decreases as the number of inward spikes increase. In particular, in the limit of infinitely many spikes, the tilt approaches to $\frac{4}{\alpha^\prime} N$.

\paragraph{Long strings.}

Finally, let us consider the long string regime. As depicted in Fig.~\ref{Fig:rho0rho1}, we always have $\rho_1<\rho_0$ even in the limit $\rho_0\to \frac{\pi}{4}$ for a finite $n$. For example, for inward spike solutions with $n=2$ and $N=1$, $\rho_1$ is bounded as $\rho_1\lesssim 0.784(<\frac{\pi}{4})$, which is saturated when $\rho_0=\frac{\pi}{4}$. Therefore, the string shape does not approach to a circular form as long as we consider a finite $n$.
This is why the upper endpoint of the Regge trajectory does not touch the vertical axis $S=0$. This is analogous to the outward spike case, but there are no analogue of rounded spike solutions that extrapolate the Regge trajectory of inward spike solutions to $S=0$, at least within our ansatz.

\bigskip
We conclude this section by summarizing implications of our results. First, in the short string regime, Regge trajectories of spiky strings have a steeper tilt than that of folded strings. This means that Regge trajectories of spiky strings are subleading Regge trajectories (whose contributions to the Regge limit amplitudes are subleading). Second, similarly to the folded string case, each Regge trajectory has the maximum spin and energy. In particular, this property is helpful for the spectra to be consistent with the Higuchi bound.  This also implies that a single Regge trajectory has a finite number of higher-spin states, in contrast to flat space and AdS. Third, we found that spiky string solutions for a fixed winding number $N$ scan a finite region on the energy-spin plane. Therefore, in order to have an infinite number of higher-spin states, we need to take into account an infinite number of Regge trajectories with an increasing winding number $N$. It would be important to further study implications of this result for high-energy scattering in de Sitter space.

\section{Spiky strings with internal motion}\label{stringinternal}
Finally, we study spiky strings with internal motion (see Ref.~\cite{Ishizeki:2008tx} for the corresponding solutions in $AdS_3\times S^1$). We employ the full ansatz~\eqref{rigid_ansatz}, under which the equations of motion are  Eqs.~\eqref{eom_rho}-\eqref{eom_phi}. For later convenience, we introduce a new variable $r$ by
\begin{align}
r = \sin^2 \rho\,, \label{r_to_rho}
\end{align}
which will be used mainly in the rest of the section instead of $\rho$. To follow the string dynamics, we first integrate the equations of motion \eqref{eom_t}-\eqref{eom_phi} as
\begin{align}
\label{internal_eom1}
C&=
\frac{\omega N r + \nu \psi^\prime}{\sqrt{\mathcal{D}}} (1-r)
\,,
\\
\label{internal_eom2}
\lambda &=\frac{\omega N r + \nu \psi^\prime}{\nu\omega \,\psi'+N(1-\nu^2 - r)} \frac{1-r}{r} \, ,
\end{align}
where $C$  and $\lambda$ are real integration constants. Notice here that nontrivial solutions with $\nu\neq0$ exist only when $\psi'$ is $\sigma$-dependent, otherwise $r$ has to be a constant. Also note that we have four parameters $(\omega,\nu,C,\lambda)$ characterizing the solutions.

\medskip
Then, we reformulate Eqs.~\eqref{internal_eom1}-\eqref{internal_eom2} such that $\psi'$ and $\rho'$ are expressed in terms of variables without derivatives. First, Eq.~\eqref{internal_eom2} implies
\begin{align}
\psi^{\prime}&=  \;N r \frac{\lambda\left(1-r-\nu^{2}\right)-\omega(1-r)}{\nu(1-r-\lambda \omega r)}\label{psi_prime} \,.
\end{align}
Second, as discussed in Appendix~\ref{subsec_app1}, we can reorganize Eq.~\eqref{internal_eom1} together with Eq.~\eqref{psi_prime} into the form,
\begin{align}
r^{\prime 2} =4r(1-r)\rho'^2=Tr^2(1-r)^2 \frac{\left(r-r_{A}\right)\left(r-r_{B}\right)\left(r -r_{C}\right)}{\left(r-r_{S}\right)^{2}}\label{rho_prime} \,.
\end{align}
This shows that for generic values of $(\omega,\nu,C,\lambda)$, $r'^2$ has a double pole and three zeros, in addition to the two double zeros located at $r=0,1$. The location of the double pole is determined by $\omega$ and $\lambda$ alone as
\begin{align}
r_S=\frac{1}{1+\lambda\omega}\,. \label{lambda}
\end{align}
On the other hand, the locations of the three zeros depend on the four parameters $(\omega,\nu,C,\lambda)$ in a more complicated manner, which we denote by $r_A$, $r_B$, and $r_C$ (see Appendix~\ref{subsec_app1} for details). Note that $r_{A,B,C}$ are complex in general. Besides, the overall constant $T$ reads
\begin{align}
T=\frac{4N^2\lambda^2(1+\omega^2)}{C^2(1+\lambda\omega)^2}\,, \label{T}
\end{align}
which is non-negative since $N$ is a positive integer and $\omega$, $\lambda$, and $C$ are real numbers. Integrating Eqs.~\eqref{psi_prime}-\eqref{rho_prime} gives string solutions for given $(\omega,\nu,C,\lambda)$.

\subsection{Outward and inward spike solutions}
Now we are ready to study shapes and Regge trajectories of the solutions described by our ansatz~\eqref{rigid_ansatz}. Our task is basically parallel to the one in Sec.~\ref{spikystrings}, but it is more complicated simply because there are more parameters of the solution. In the present paper, for illustration, we focus on
two classes of solutions
that reduce to those of the previous section in the limit $J\to0$, which simplifies the analysis considerably.
We call them outward spike solutions and inward spike solutions by analogy with the solutions in Sec.~\ref{spikystrings}. In the following, we present properties of these solutions.

\paragraph{Ansatz on $r_A$, $r_B$, $r_C$, and $r_S$.}

In Sec.~\ref{spikystrings}, we demonstrated that shapes of the string depend on the location of zeros and poles of $\rho'^2$. Similarly, the outward and inward spike solutions can be classified based on the values of $r_A$, $r_B$, $r_C$, and $r_S$. First, for both classes of solutions, $r_A$, $r_B$, and $r_C$ are all real and positive. Without loss of generality, we assume that $r_A < r_B < r_C$. These values relative to $r_S$ are also relevant for us, based on which we perform the following classification:
\begin{align}
\text{Outward spike solutions: } &r_A < r_B < r_S < r_C\,, \label{outansatz}
\\
\text{Inward spike solutions: }&r_S < r_A < r_B < r_C\,. \label{inansatz}
\end{align} 
In Appendix~\ref{subsec_app2}, we show that in the limit $J \rightarrow 0$, these solutions indeed reduce to their counterparts in Sec.~\ref{spikystrings}.

\paragraph{Reality conditions.}

Next let us take care of reality conditions. First, Eq.~\eqref{rho_prime} shows that reality of $r(\sigma)$ requires $r_A\leq r(\sigma)\leq r_B$ or $r(\sigma)\geq r_C$ (recall that the overall coefficient $T$ is positive). Also, in order for the closed string to form a loop, $r'$ has to flip the sign somewhere on the worldsheet, otherwise the string stretches forever. Then, for the outward and inward spike solutions, the string has to be inside the regime $r_A\leq r(\sigma)\leq r_B$.

 \begin{figure}[t] 
	\centering 
	\includegraphics[width=6.7cm]{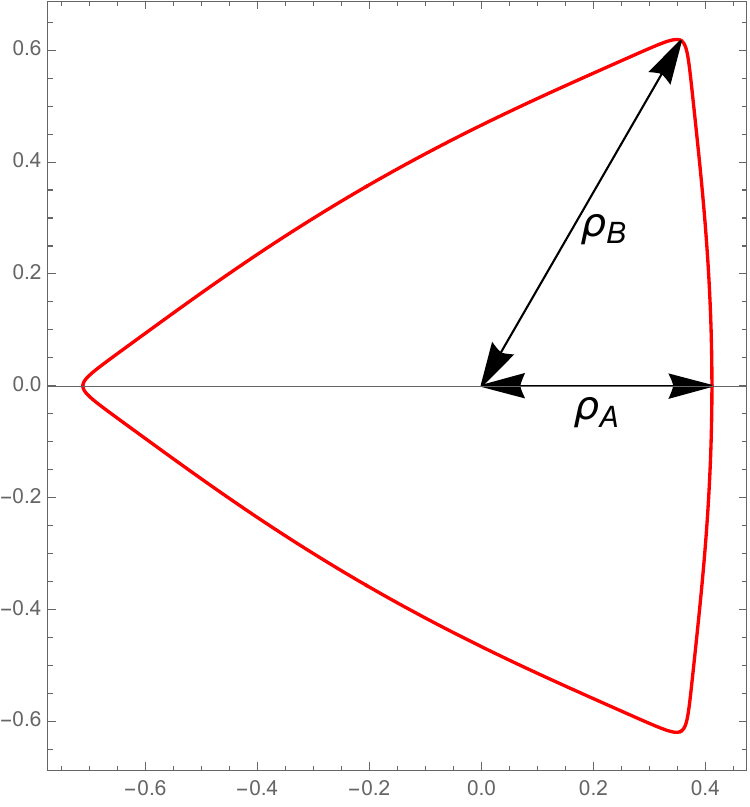}    \quad  \quad  \quad
	\includegraphics[width=6.7cm]{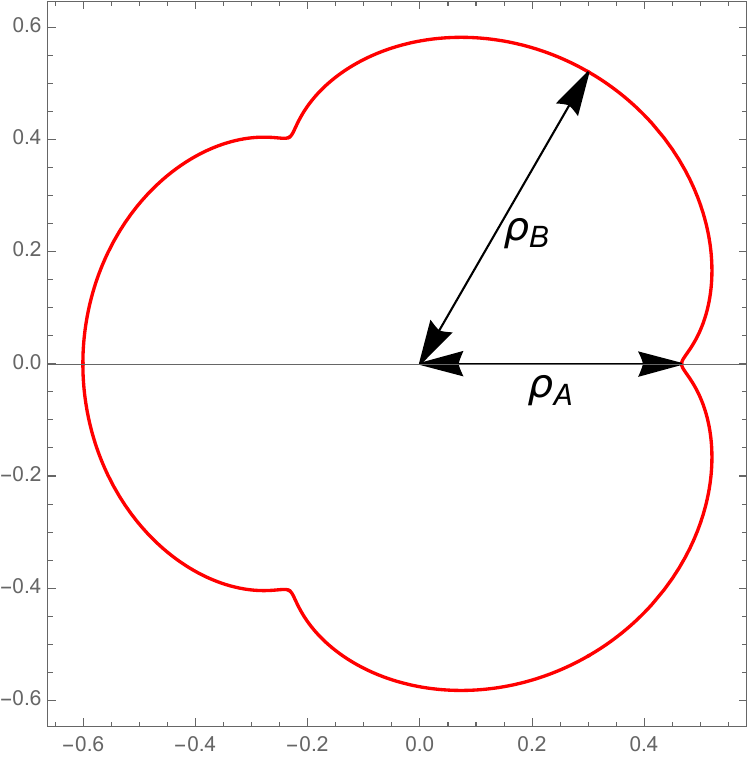}  
	\caption{Shapes of outward spike solutions and inward spike solutions: $\rho_A$ and $\rho_B$ are defined by
$r_A=\sin^2\rho_A$ ($0\leq \rho_A\leq\frac{\pi}{2}$) and similarly for $\rho_B$. For outward spike solutions, we chose $(\omega,\nu,C,\lambda) \simeq (0.89,0.47,0.41,1.31)$, which corresponds to $(\rho_{A},\rho_{B}) \simeq (0.41,0.71)$. For inward spike solutions, we chose $(\omega,\nu,C,\lambda) \simeq (1.65,0.51,0.94,2.57)$, which corresponds to $(\rho_{A},\rho_{B}) \simeq (0.47,0.60)$.}
 \label{shape_internal}
 \end{figure} 

\paragraph{Periodicity conditions.}

Finally, we take into account global structures of the string. As before, the angle $\Delta\phi$ (on the $r$-$\phi$ plane) between the two spikes\footnote{
As we see shortly, the string is smooth everywhere for $\nu\neq0$, but we can interpret that spikes are rounded, similarly to the rounded spikes in Sec.~\ref{spikystrings}. Therefore, we use the terminology ``spikes" as before.} has to be quantized appropriately. More explicitly, for $n$-spike solutions, we require
\begin{align}
\Delta \phi=\frac{2\pi N}{n}\,.
\end{align}
Within the ansatz \eqref{outansatz}-\eqref{inansatz}, an explicit form of $\Delta\phi$ reads
\begin{align}
\Delta\phi=2N\int_{r_A}^{r_B}\frac{dr}{r'}
=\frac{2N}{\sqrt{T}}\int_{r_A}^{r_B}
\frac{dr}{r(1-r)} \frac{|r-r_{S}|}{\sqrt{\left(r-r_{A}\right)\left(r-r_{B}\right)\left(r -r_{C}\right)}}\,.
\end{align}
In the present setup, we also need to take care of periodicity along the internal $S^1$. For simplicity, we assume that the string has no winding along the $S^1$, which implies
\begin{align}
0=\int_0^{2\pi}d\sigma \psi'=2n\int_{r_A}^{r_B}\frac{dr}{r'}\psi'
=\pm\frac{2nNr_S}{\nu\sqrt{T}}\int_{r_A}^{r_B}\frac{dr}{1-r} \frac{\lambda\left(1-\nu^{2}-r\right)-\omega(1-r)}{\sqrt{\left(r-r_{A}\right)\left(r-r_{B}\right)\left(r -r_{C}\right)}}\,.
\end{align}
Here the plus and minus signs are for outward and inward spike solutions, respectively. As we mentioned, there are four parameters of the solutions. If we specify the number of spikes $n$ and the winding number $N$, there are two constraints originating from the periodicity conditions. Then, we are left with two degrees of freedom characterizing the size of the string and the internal motion.

\paragraph{Shapes.}

In Fig.~\ref{shape_internal}, we illustrate outward and inward spike solutions for $n=3$ and $N=1$. The four parameters $(\omega,\nu,C,\lambda)$ are chosen such that the two periodicity conditions are satisfied. In contrast to the case without internal motion, the spikes are indeed rounded.

\paragraph{Regge trajectories.}

Finally, we study Regge trajectories. First, substituting Eqs.~\eqref{internal_eom1}-\eqref{internal_eom2} into Eqs.~\eqref{E_master}-\eqref{J_master}, we find a simplified expression for conserved charges\footnote{
To derive them, it is convenient to use Eq.~\eqref{convenient1} and Eq.~\eqref{D1} provided in Appendix.}:
\begin{align}
E &=  \frac{NR}{2\pi\alpha'}{ \frac{\lambda}{C}} \int_{0}^{2 \pi} d \sigma \, r\, \frac{(1-r)^{2}-C^{2}}{1-r-\lambda \omega r} \\*
&= \pm   \frac{NR}{2\pi\alpha'} \frac{2n \lambda r_S}{C\sqrt{T}}\int_{r_{A}}^{r_{B}} \frac{d r}{1-r} \frac{(1-r)^2-C^2}{\sqrt{\left(r-r_{A}\right)\left(r-r_{B}\right)\left(r-r_{C}\right)}}\,,  \\
S&= \frac{N R^2}{2\pi \alpha' }{\frac{1}{C}}\int_{0}^{2 \pi} d \sigma \, r \, \frac{(1-r) r \lambda \omega-C^{2}}{1-r-\lambda \omega r}\,,\\*
&= \pm   \frac{NR^2}{2\pi\alpha'} \frac{2n r_S}{C\sqrt{T}}\int_{r_{A}}^{r_{B}} \frac{d r}{1-r} \frac{(1-r)r \lambda \omega-C^2}{\sqrt{\left(r-r_{A}\right)\left(r-r_{B}\right)\left(r-r_{C}\right)}}\,,  \\
J&= \frac{N R^2}{2\pi\alpha'} \frac{1}{ \nu C } \int_{0}^{2 \pi}d \sigma \,r \, \frac{(1-r) \lambda \nu^{2}-C^{2}(\lambda-\omega)}{1- r -\lambda \omega r}\\*
&= \pm   \frac{NR^{2}}{2\pi\alpha'} \frac{2n r_S}{\nu C\sqrt{T}}\int_{r_{A}}^{r_{B}} \frac{d r}{1-r} \frac{(1-r) \lambda \nu^2 - C^2(\lambda - \omega)}{\sqrt{\left(r-r_{A}\right)\left(r-r_{B}\right)\left(r-r_{C}\right)}}\,,
\label{J_last}
\end{align}
where the plus and minus signs are again for outward and inward spikes, respectively.

\begin{figure}[t] 
	\centering 
	\includegraphics[width=7.3cm]{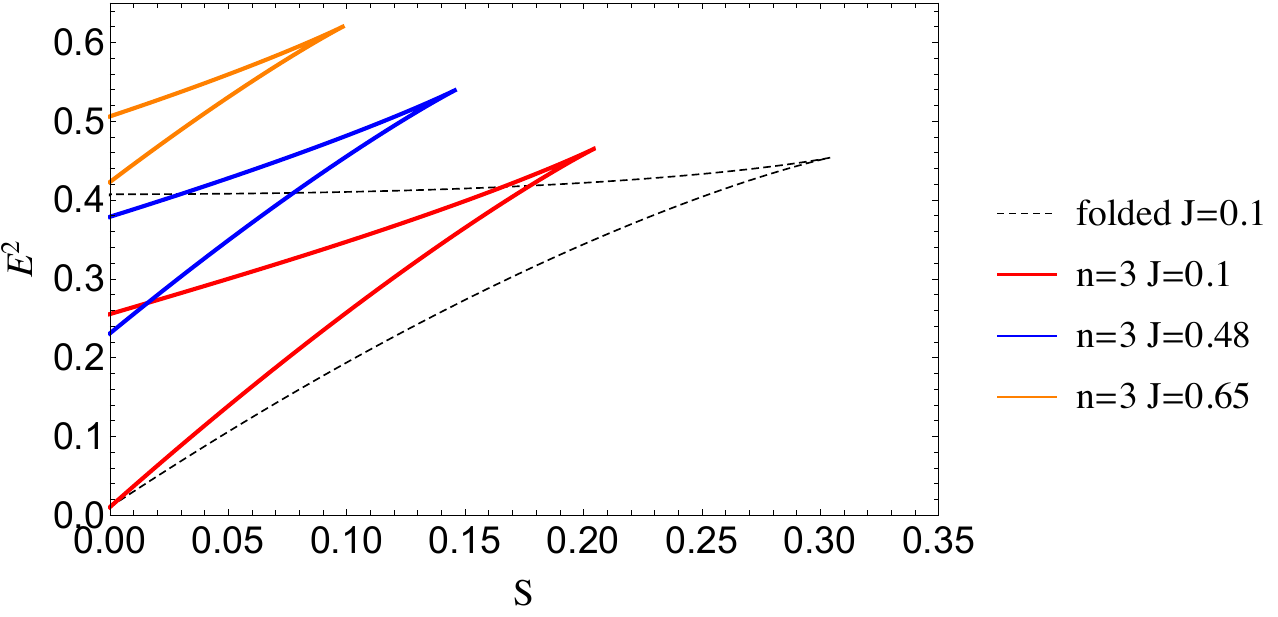}    \quad  
	\includegraphics[width=7.3cm]{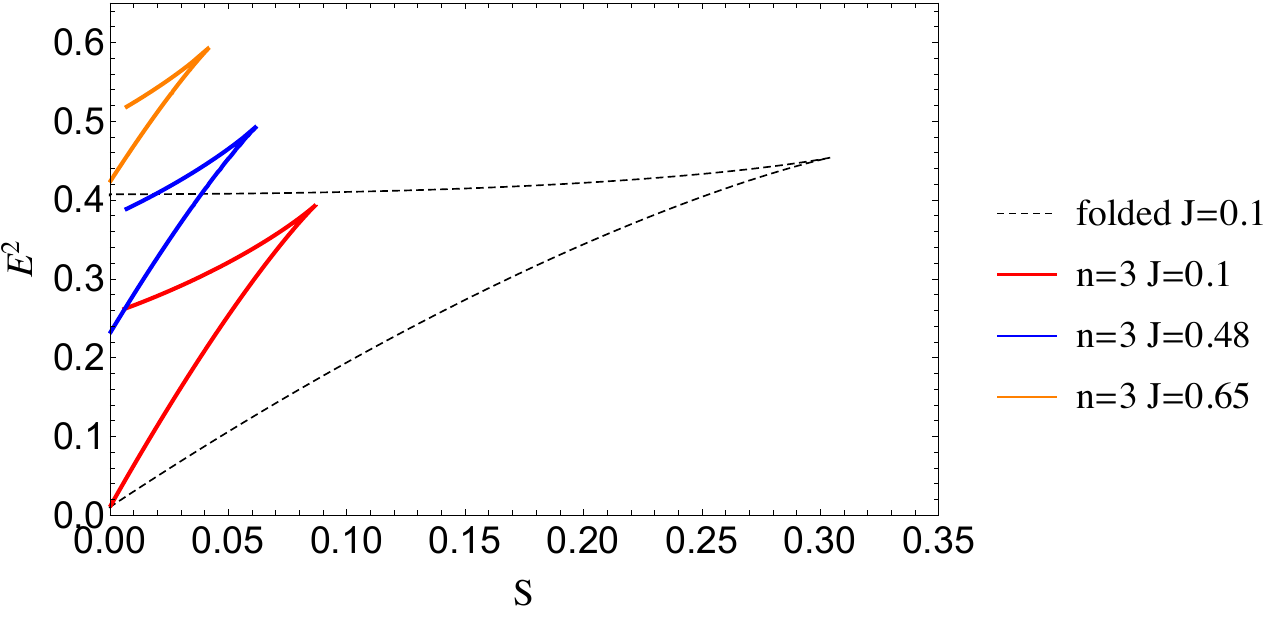}  
	\caption{Regge trajectories of spiky strings with internal charges. The left and right figures are for outward and inward spike solutions, respectively.  For comparison, we also illustrate the Regge trajectory of the folded string in the dotted lines. The energy, spin and internal charge are in the units of  $R/{\alpha^\prime}$, $R^2/{\alpha^\prime}$ and $R^2/{\alpha^\prime}$, respectively. } \label{InternaldSRegge}
\end{figure}

\medskip
As we mentioned, once we specify the number of spikes $n$ and the winding number $N$, we are left with two degrees of freedom associated with the size of the string and the internal motion. If we further specify the internal charge through Eq.~\eqref{J_last}, we are left with one degree of freedom characterizing the size of the string. Then, by varying the size of the string, we can draw Regge trajectories for fixed $n$, $N$ and $J$. See Fig.~\ref{InternaldSRegge} for Regge trajectories of outward and inward spike solutions with $n=3$, $N=1$, and different values of $J$. We find that as the internal charge increases, the Regge trajectory shifts upwards. Also, the maximum spin decreases and the maximum energy increases. In particular, Regge trajectories for fixed $n$ and $N$ scan a finite region of the energy-spin plane. These properties are qualitatively the same as folded strings with internal charges and spiky strings without internal charges, respectively. Besides, we find that Regge trajectories for outward spikes touch the vertical axis $S=0$ twice. This explains that in the limit $J\to0$, outward spike solutions reduce to both the outward and rounded spike solutions presented in the previous section. See Appendix~\ref{subsec_app2} for more details.

\section{Summary and discussion}\label{conclusion}

In this paper, we studied a class of semiclassical strings in de Sitter space and the corresponding Regge trajectories. Within the rigid string ansatz~\eqref{rigid_ansatz}, there are two classes of solutions: folded strings and spiky strings. First, we showed for folded strings that there exist the maximum spin and energy in each Regge trajectory for a fixed internal charge and a fixed folding number. This means that a single Regge trajectory includes only a finite number of higher spin states in sharp contrast to the flat space and AdS ones. Also, as the internal charge increases, the maximum spin decreases. While this property is helpful for the spectrum to be consistent with the Higuchi bound, it implies that Regge trajectories with a fixed folding number (and different internal charges) scan a finite region of the energy-spin plane. We demonstrated that the same properties hold for spiky strings too. This implies that in order to have infinitely many higher-spin states (within our ansatz), one needs to consider infinitely many Regge trajectories with an increasing folding number.

\medskip
More intuitively, the above mentioned properties are natural consequences of de Sitter acceleration. First, the string can have a large spin if it is long and rotates with a large angular velocity. On the other hand, causality requires that the string worldsheet cannot propagate faster than the speed of light, which gives an upper bound on the string length in terms of the angular velocity. In flat space and AdS, the string stretches with an infinite length if the angular velocity approaches to zero. In particular, the large string length competes against the smallness of the angular velocity, so that strings have larger spins as they spread more. In sharp contrast, de Sitter space has an acceleration, so that there exists a natural cutoff dictated by causality: the string cannot rotate anymore when touching the horizon. Therefore, the only way for a string to have a large spin is to shrink inside the horizon, fold as much as possible, and rotate quickly. This is why string Regge trajectories in de Sitter space are qualitatively different from the flat space and AdS ones. Besides, de Sitter acceleration makes  spiky strings fatter, leading to several new classes of solutions which do not exist in flat space and AdS.

\medskip
As a concluding remark, we would like to present several interesting future directions. The first question to ask is about high-energy behavior of string scattering in de Sitter space. Recall that in flat space and AdS, string scattering has a mild high-energy behavior due to existence of infinitely many higher-spin states. In particular, high-energy scattering is captured by a widely spreading worldsheet, which implies an exponential suppression of the amplitudes. On the other hand, in order to have sufficiently many higher-spin states in de Sitter space, one needs to consider strings shrinking and folding inside the horizon. At least naively, this would suggest that the way of stringy UV completion could be different in de Sitter space compared to flat space and AdS. It would be important to study this issue further by generalizing developments in holographic correlation functions in AdS~\cite{Kazama:2016cfl,Kazama:2013qsa,Caetano:2012ac,Kazama:2011cp,Janik:2011bd,Costa:2010rz,Zarembo:2010rr}, which would provide cosmological Veneziano amplitudes. A related important question is to formulate a framework to study consistency of high-energy scattering in de Sitter space. For example, in the case of AdS, we know what are the AdS analogues of the Regge limit amplitudes and the hard scattering amplitudes (see, e.g.,~\cite{Gary:2009ae,Okuda:2010ym,Penedones:2010ue,Raju:2012zr, Costa:2012cb, Maldacena:2015iua, Haldar:2019prg,Meltzer:2019pyl,Komatsu:2020sag,Chandorkar:2021viw}). For de Sitter space, there is a known flat space limit of late-time correlators corresponding to the hard scattering limit (see, e.g.,~\cite{Raju:2012zr, Arkani-Hamed:2015bza,Arkani-Hamed:2017fdk}). However, to our knowledge, its understanding is still limited compared to the AdS case, even at the quantum field theory level before taking into account stringy effects. It would be important to clarify which kinematics of which quantities is useful to define consistency of high-energy scattering in de Sitter space. We hope that this direction would open up a new road toward understanding of de Sitter space in string theory.

\section*{Acknowledgements}
We are deeply grateful to Shinji Hirano for fruitful discussion and encouragement since the early stage of this work. We would like to thank Guilherme Pimentel for the chance of presenting this work in the online workshop ``Cosmological Correlators" and many helpful discussions during the workshop. We also thank Pablo Soler and Bo Sundborg for useful discussions. M.K. is supported in part by JSPS KAKENHI Grant Number 20K03966. T.N. is supported in part by JSPS KAKENHI Grant Numbers JP17H02894 and 20H01902. T.T. is supported in part by the Iwanami Fujukai Foundation. S.Z. is supported in part by the Swedish Research Council under grant numbers 2015-05333 and 2018-03803.

\appendix
\section{Details of spiky strings with internal motion}
\label{appendix}

In this appendix, we summarize details of spiky strings with internal motion.

\subsection{Derivation of Eq.~\eqref{rho_prime}}
\label{subsec_app1}

We begin by providing a derivation of Eq.~\eqref{rho_prime}. For this, it is convenient to note the following relation which follows from Eq.~\eqref{psi_prime}:
\begin{align}
\label{convenient1}
\omega Nr+\nu\psi'=N\lambda r_Sr\frac{(1-\nu^2)-(1+\omega^2)r}{r_S-r}\,,
\end{align}
where we defined
\begin{align}
r_S=\frac{1}{1+\lambda\omega}\,.
\end{align} 
Substituting this into Eq.~\eqref{internal_eom1} gives
\begin{align}
\mathcal{D}&=\frac{(\omega Nr+\nu\psi')^2(1-r)^2}{C^2}
=\frac{N^2\lambda^2r_S^2}{C^2}\cdot\frac{r^2(1-r)^2}{(r_S-r)^2}\cdot \left((1-\nu^2)-(1+\omega^2)r\right)^2\,.
\label{D1}
\end{align}
On the other hand, we can reformulate Eq.~\eqref{D} using Eq.~\eqref{psi_prime} as
\begin{align}
\mathcal{D}
&=\left((1-\nu^2)-(1+\omega^2)r\right)
\left(\rho'^2-\frac{N^2r_S^2}{\nu^2}\cdot\frac{r(1-r)}{(r_S-r)^2}\cdot F(r)\right)
\,,
\label{D2}
\end{align}
where $F(r)$ is a quadratic polynomial defined by
\begin{align}
\label{F_r}
F(r)=(\lambda-\omega)^2r^2+\left((1+\lambda^2)\nu^2-(\lambda-\omega)^2\right)r-\nu^2\,.
\end{align}
Comparing Eq.~\eqref{D1} and Eq.~\eqref{D2} gives
\begin{align}
\rho'^2=\frac{N^2\lambda^2r_S^2}{C^2}\cdot\frac{r(1-r)}{(r_S-r)^2}\cdot \left[r(r-1)\left((1+\omega^2)r-(1-\nu^2)\right)+\frac{C^2}{\lambda^2\nu^2}F(r)\right]\,.
\end{align}
Then, we conclude that
\begin{align}
r'^2=4r(1-r)\rho'^2
=\frac{4N^2\lambda^2(1+\omega^2)r_S^2}{C^2}\cdot\frac{r^2(1-r)^2}{(r_S-r)^2}\cdot 
\left[r^3+\widetilde{F}(r)\right]\,,
\end{align}
where $\widetilde{F}(r)$ is a quadratic polynomial defined by
\begin{align}
\label{Ftilde_r}
\widetilde{F}(r)=\frac{1}{1+\omega^2}
\left[
-\big((1+\omega^2)+(1-\nu^2)\big)r^2+(1-\nu^2)r+\frac{C^2}{\lambda^2\nu^2}F(r)
\right]\,.
\end{align}
This reproduces Eq.~\eqref{rho_prime} by identifying $r_{A,B,C}$ with three solutions for $r^3+\widetilde{F}(r)=0$.

\subsection{$J=0$ limit}
\label{subsec_app2}

\begin{figure}[t] 
	\centering 
	\includegraphics[width=15.2 cm]{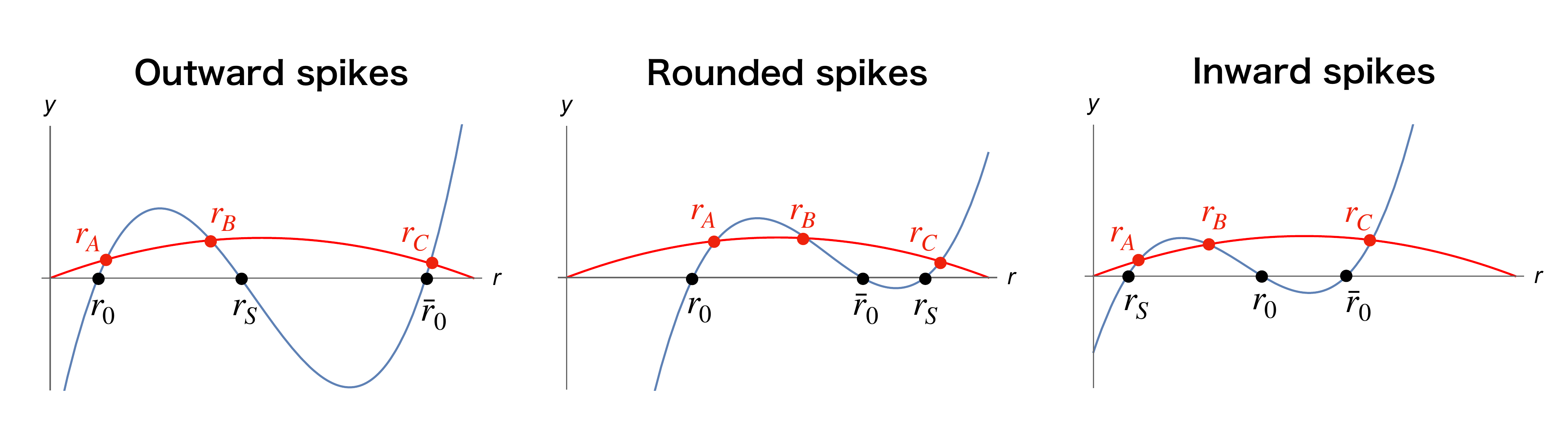} 
	\caption{The blue curves are $y=(r-r_S)\left(r^2-r+\frac{C^2}{\omega^2}\right)$, which intersect with the $r$-axis at $r= r_0,\bar{r}_0,r_S$. The red curves are $y = \nu^2r_Sr(1-r)$. The intersection points of the blue and red curves are the three solutions $r_{A,B,C}$ of Eq.~\eqref{defeqABClimit}. In the limit $\nu \rightarrow 0$, the three solutions approach to $r_0,\bar{r}_0, r_{S}$, and the solutions in Sec.~\ref{spikystrings} are reproduced. For example, when $r_0<r_S<\bar{r}_0$, finite $\nu$ solutions with $r_A<r_B<r_S<r_C$ are reduced to the outward spike solutions.
} \label{no_internal_limit}
\end{figure} 

Finally, we discuss the limit where the internal charge vanishes $J=0$. First, the internal velocity $\nu$ and the internal space dependence $\psi'$ of the string have to vanish to reproduce the solutions in Sec.~\ref{spikystrings}. In particular, Eq.~\eqref{psi_prime} shows that this is achieved in the limit
\begin{align}
(\lambda-\omega)^2\ll\nu^2\ll1\,.
\end{align}
Note that $\psi'$ diverges if we take the limit $\nu^2\ll(\lambda-\omega)^2\ll1$ instead. Then, let us study properties of $r_{A,B,C}$ for $\lambda=\omega$ with a finite $\nu$. Under this assumption, the defining equation $r^3+\widetilde{F}(r)=0$ of $r_{A,B,C}$ is reduced to
\begin{align}
(r-r_S)\left(r^2-r+\frac{C^2}{\omega^2}\right)
=\nu^2r_Sr(1-r)\,, \label{defeqABClimit}
\end{align}
where note that $r_S=(1+\omega^2)^{-1}$ in the limit $\lambda=\omega$. If we further take the limit $\nu\to0$, one of $r_{A,B,C}$ coincides with $r_S$. Therefore, the double pole at $r=r_S$ and one of three zeros collide and form a single pole at $r=r_S$, which is identified with a single pole of $\rho'^2$ at $\rho=\rho_1$ shown in Eq.~\eqref{spikyansatz}.

\medskip
To see how the limit $\nu\to0$ reproduces the three classes of solutions in Sec.~\ref{spikystrings}, let us parameterize the two solutions  $r=r_0,\bar{r}_0$ for $r^2-r+\frac{C^2}{\omega^2}=0$ as
\begin{align}
r_0=\sin^2\rho_0\,,
\quad
\bar{r}_0=\sin^2(\tfrac{\pi}{2}-\rho_0)=\cos^2\rho_0\,,
\end{align}
 where $\rho_0$ is identified with that in Sec.~\ref{spikystrings}. Notice that we can employ this parameterization without loss of generality since $r_0+\bar{r}_0=1$. Also,  in order for $r_0$ and $\bar{r}_0$ to be real, $0\leq  r_0  \bar{r}_0  = \frac{C^2}{\omega^2}\leq \frac{1}{4}$ has to be satisfied, under which  we can choose $\rho_0$ such that $
 0\leq \rho_0\leq\frac{\pi}{2}$.
 The classification in Sec.~\ref{spikystrings} is then rephrased as
\begin{itemize}
\item $r_0<r_S<\bar{r}_0$: outward spike solutions,
\item $r_0<\bar{r}_0<r_S$: rounded spike solutions,
\item $r_S<r_0<\bar{r}_0$: internal spike solutions.
\end{itemize}
As depicted in Fig.~\ref{no_internal_limit}, the outward spike solutions and rounded spike solutions are obtained in the limit $\nu\to0$ of solutions with the ordering $r_A<r_B<r_S<r_C$, whereas the internal spike solutions are obtained from those with $r_S<r_A<r_B<r_C$.

\bibliography{dS_string}{}
\bibliographystyle{utphys}

\end{document}